\documentclass[twocolumn,aps,pre,showpacs,superscriptaddress]{revtex4}
\usepackage[utf8]{inputenc}
\usepackage[T1]{fontenc}
\usepackage{graphicx}% Include figure files
\usepackage{amsmath,float}
\begin{document}
\title{Arbitrary amplitude nucleus-acoustic solitary waves in thermally degenerate plasma systems}
\author{A. Mannan}
\email{abdulmannan@juniv.edu}
\affiliation{Department of Physics, Jahangirnagar University, Savar, Dhaka-1342, Bangladesh}
\affiliation{Institut f\"{u}r Mathematik, Martin Luther Universit\"{a}t Halle-Wittenberg, D-06099 Halle (Saale), Germany}
\author{S. Sultana}
\affiliation{Department of Physics, Jahangirnagar University, Savar, Dhaka-1342, Bangladesh}
\author{A. A. Mamun}
\altaffiliation[Also at~]{Wazed Miah Science Research Centre, Jahangirnagar University, Savar, Dhaka-1342, Bangladesh.}
\affiliation{Department of Physics, Jahangirnagar University, Savar, Dhaka-1342, Bangladesh}
\date{\today}
\begin{abstract}
A rigorous theoretical investigation is made of arbitrary amplitude nucleus acoustic solitary waves in a fully ionized multi-nucleus plasma system (consisting of thermally degenerate electron species and non-degenerate warm light as well as heavy nucleus species). The pseudo-potential approach, which is valid for the arbitrary amplitude solitary waves, is employed. The subsonic and supersonic nucleus-acoustic solitary waves (which are found to be compressive) along with their basic features are identified. The basic properties  of these subsonic and supersonic nucleus-acoustic solitary waves are found to be significantly modified by the effects of non and ultra-relativistically degenerate electron species, dynamics of heavy nucleus species, number densities as well as adiabatic temperatures of light and heavy nucleus species, etc. It shown that the presence of heavy nucleus species with non-degenerate (isothermal) electron species supports the existence of subsonic nucleus-acoustic solitary waves, and that the effects of electron degeneracies and light and heavy nucleus temperatures reduce the possibility for the formation of these subsonic nucleus-acoustic solitary waves. The amplitude of the supersonic nucleus-acoustic solitary waves in the situation of non-relativistically degenerate electron species is much smaller than that of ultra-relativistically degenerate electron species, but is much larger than that of isothermal electron species. The rise of adiabatic temperature of light or heavy nucleus species causes to decrease (increase) the amplitude (width) of the subsonic and supersonic nucleus acoustic solitary waves. On the other hand, the increase in the number density of light or heavy nucleus species causes to increase (decrease) the amplitude (width) of the subsonic and supersonic nucleus acoustic solitary waves. The results of this investigation are found to be applicable in laboratory, space, and astrophysical plasma systems.
\end{abstract}
\maketitle
\section{\label{Intro}Introduction}
Mamun \cite{Mamun17-18} has first introduced the electron degenerate energy along with corresponding wave speed ($C_l$) and wave scale length ($L_q$) associated with the degenerate electron pressure \cite{Chandrasekhar31,Horn91,Fowler94,Koester02,Shukla11a,Shukla11b,Brodin16}, and has also  identified the degenerate pressure driven (DPD) nucleus-acoustic (NA) waves, and has pinpointed their new basic features in degenerate plasma systems  \cite{Chandrasekhar31,Horn91,Fowler94,Koester02,Shukla11a,Shukla11b,Brodin16,Killian2006,Fletcher2006,Glenzer2009}, which are composed of cold degenerate electron species (DES) \cite{Chandrasekhar31,Horn91,Fowler94}, cold non-degenerate light nucleus species
(viz. ${\rm ~^{1}_{1}H}$ \cite{Chandrasekhar31}, or ${\rm ~^{4}_{2}He}$ \cite{Horn91} or ${\rm ~^{12}_{~6}C}$ \cite{Koester02} or ${\rm ~^{16}_{~8}O}$ \cite{Koester02}), and stationary heavy nucleus species (viz. ${\rm ~^{56}_{26}Fe}$ \cite{Vanderburg15} or ${\rm ~^{85}_{37}Rb}$ \cite{Witze14} or
${\rm ~^{96}_{42}Mo}$ \cite{Witze14}). The linear dispersion relation for such DPD NA waves in such a cold degenerate plasma is given by \cite{Mamun17-18}
\begin{eqnarray}
\omega=\sqrt{\frac{\gamma_e}{1+\mu}}\frac{k C_l}{\sqrt{1+\frac{\gamma_e}{1+\mu} k^2L_q^2}},
\label{NA-disp}
\end{eqnarray}
where $\omega=2\pi f$ and $k=2\pi/\lambda$ with $f$ ($\lambda$) being the DPD NA wave frequency (wavelength);
$\mu=Z_hN_{h0}/Z_lN_{l0}$ with $Z_le$ ($Z_he$) being the charge of the light (heavy) nucleus species,
and $N_{l0}$ ($N_{h0}$) being the equilibrium number density of the light (heavy) nucleus species; $C_l= (Z_lE_{e0}/m_l)^{1/2}$ is the DPD NA wave speed with $m_l$ being the mass of a light nucleus; $E_{e0}=KN_{e0}^{\gamma_e-1}$ is the cold degenerate electron energy \cite{Mamun17-18} associated with degenerate electron pressure \cite{Chandrasekhar31,Horn91,Fowler94} $P_{e0}=K{N_{e0}}^{\gamma_e}$ at equilibrium; $K\simeq3\pi\hbar^2/5m_e$ \cite{Mamun17-18,Chandrasekhar31,Horn91,Fowler94,Koester02,Shukla11a,Shukla11b} for
$\gamma_e=5/3$ (non-relativistically DES \cite{Chandrasekhar31,Horn91,Fowler94}); $K\simeq 3\hbar c/4$ \cite{Mamun17-18,Chandrasekhar31,Horn91,Fowler94,Koester02,Shukla11a,Shukla11b} for $\gamma_e=4/3$ (ultra-relativistically DES \cite{Chandrasekhar31,Horn91,Fowler94});
$L_q=C_l/\omega_{pl}$ is the DPD NA wave length scale with $\omega_{pl}= (4\pi N_{l0}Z_l^2e^2/m_l)^{1/2}$ being the nucleus plasma frequency and $m_l$ mass of a light nucleus; $m_e$ ($\hbar$) is the electron rest mass (reduced Planck's constant), $c$ is the speed of light in vacuum, and $e$ is the charge of a proton or the magnitude of the charge of an electron. We note that $N_{e0}=Z_lN_{l0}+Z_hN_{h0}$ at equilibrium.
It is important to mention that in any cold degenerate plasma $K$ is unknown for $\gamma_e=1$, which, thus, cannot be considered in (\ref{NA-disp}) since the latter is not valid for any cold degenerate plasma system. The dispersion relation for the long wavelength DPD NA waves (viz. $kL_q\ll 1$, which is the appropriate limit for these waves) becomes
\begin{equation}
\omega\simeq \sqrt{\frac{\gamma_e}{1+\mu}}\, kC_l,
\label{NA-Ldisp}
\end{equation}
which indicates that the degenerate electron pressure (nucleus mass density) provides the restoring force (inertia) in these DPD NA waves, and that  the phase speed of these DPD NA waves decreases (increases) with the rise of $\mu$ ($\gamma_e$). This dispersion relation also indicate that the DPD NA waves completely disappear in absence of the electron degenerate pressure,
which is independent of the temperature of any plasma species. Thus, the DPD NA waves \cite{Mamun17-18} defined by (\ref{NA-disp}) is completely different from the well known ion-acoustic (IA) waves \cite{Tonks29,Revans33,Buti68} due to the fact that the IA (DPD NA) waves are driven by the electron thermal (degenerate) pressure, and that the speed, length scale, and time scale of the IA waves are far different from those of the DPD NA waves. However, cold degenerate plasma systems under different conditions have been considered by many authors to study the nonlinear propagation of the IA waves during the last ten years \cite{Mamun10a,Mamun10b,El-Labany10,Hossain11,Misra11,Akhtar11,Roy12,Haider12,Nahar13,Zobaer13a,Zobaer13b,El-Labany14,Hossen14,Rahman15,Hossen15,El-Labany16,Hosen17,Hasan17,Islam17a,Islam17b,Sultana18a,Hosen18,Patidar20}.

Recently, there has been a great deal of interest in understanding the physics of linear and nonlinear propagation of DPD NA waves \cite{Mamun17-18}  in degenerate plasma systems under different situations  \cite{Chandrasekhar31,Horn91,Fowler94,Koester02,Shukla11a,Shukla11b,Brodin16,Killian2006,Fletcher2006,Glenzer2009,Vanderburg15,Witze14} not only because of their basic difference from the IA waves \cite{Tonks29,Revans33,Buti68}, but also because of the existence of the degenerate plasma systems \cite{Chandrasekhar31,Horn91,Fowler94,Koester02,Shukla11a,Shukla11b,Brodin16,Killian2006,Fletcher2006,Glenzer2009,Vanderburg15,Witze14}  in enormous number of astrophysical compact objects \cite{Chandrasekhar31,Horn91,Fowler94,Koester02,Shukla11a,Shukla11b,Brodin16} and laboratory devices \cite{Killian2006,Fletcher2006,Glenzer2009}, where the degenerate pressure is comparable to or greater than all other pressures like thermal, electrostatic, and self-gravitational pressures
\cite{Chandrasekhar31,Horn91,Fowler94,Koester02,Shukla11a,Shukla11b,Brodin16,Killian2006,Fletcher2006,Glenzer2009,Vanderburg15,Witze14}.

We are now interested in deriving a more general and realistic dispersion relation by considering thermally DES (TDES) [instead of the cold degenerate DES considered in (\ref{NA-disp})], and cold mobile heavy nucleus species [instead of stationary heavy nucleus species
considered in (\ref{NA-disp})]. The dynamics of the light nucleus species is as before. The dispersion relation for the DPD NA waves in such  thermally degenerate plasma system (TDPS) is given by
\begin{equation}
 \omega = \sqrt{\frac{\gamma_e(1+\mu S_h)}{1+\mu+\gamma _ek^2\lambda_q^2}}kC_q\,,
 \label{NA-mdisp}
\end{equation}
where $S_h = Z_hm_l/Z_lm_h$; $C_q=(Z_l\mathcal{E}_{e0}/m_l)^{1/2}$ in which $\mathcal{E}_{e0} = \mathcal{E}_{ed} + \mathcal{E}_{et}$ with
$\mathcal{E}_{ed}$ ($\mathcal{E}_{et}$) being the electron degenerate (thermal) energy associated with electron degenerate (thermal) pressure; and $\lambda_q=C_q/\omega_{pl}$.
The dispersion relation for the long-wavelength DPD NA waves ($k\lambda_q\ll 1$, which is the appropriate limit for these waves) in a TDPS becomes
 \begin{equation}
 \omega \simeq \sqrt{\frac{\gamma_e(1+\mu S_h)}{1+\mu}}kC_q\,,
 \label{NA-Lmdisp}
\end{equation}
which indicates that the dispersion relation (\ref{NA-Lmdisp}) for the DPD NA waves in such a TDPS can be interpreted as follows
\begin{itemize}
 \item{The dispersion relations (\ref{NA-Ldisp}) and (\ref{NA-Lmdisp}) are identical for $S_h=0$ (indicating stationary heavy nucleus species) and  $\mathcal{E}_{et}=0$ (indicating cold DES).}

\item{The phase speed ($\omega/k$) of the DPD NA waves increases with rise of the value of $S_h$. The rate of increase of $\omega/k$ with $\mu$ in the case of $S_h\ne 0$ is slower than that in the case of $S_h=0$.}

\item{It is obvious that $C_q>C_l$ and $\lambda_q>L_q$. This means that the phase speed (wavelength) for $\mathcal{E}_{et} \ne 0$ is higher (lower) than that for $\mathcal{E}_{et}= 0$. This is due to the rise of the volume of the degenerate medium caused by the outward thermal pressure of the TDES.}
\end{itemize}
There are also a number of investigations \cite{Mamun16,Mamun17,Jannat18a,Jannat18b,Sultana18b,Sultana18c,Chowdhury18,Zaman18,Das19} on nonlinear NA waves in degenerate plasma systems during the last five years. The limitations of these works are as follows.
\begin{itemize}
 \item{The works \cite{Mamun16,Mamun17,Jannat18a,Jannat18b,Sultana18b,Sultana18c,Chowdhury18,Zaman18,Das19} are valid only for cold degenerate electron and nucleus species. So the works are not valid for warm degenerate plasma systems, particularly for hot white dwarfs \cite{Dufour08,Dufour11,Werner15,Werner19,Koester20}.}

 \item{The works \cite{Mamun16,Mamun17,Jannat18a,Jannat18b,Sultana18b,Sultana18c,Chowdhury18,Zaman18,Das19} are based on the reductive perturbation method \cite{Washimi66}, which is valid for small amplitude nonlinear waves. Thus, the works are not valid for large amplitude nonlinear waves.}

\item{The wave speed and length scale, which are independent of the degenerate electron pressure, are not properly defined in the works \cite{Mamun16,Mamun17,Jannat18a,Jannat18b,Sultana18b,Sultana18c,Chowdhury18,Zaman18,Das19} from which one cannot get the linear dispersion relation for the DPD NA waves defined by (\ref{NA-Ldisp}) or (\ref{NA-Lmdisp}). Therefore, the linear and nonlinear features of the DPD NA waves were not properly identified by these works, which are correct for other kind of nonlinear NA waves, but not for the DPD NA waves defined by (\ref{NA-Ldisp}) or (\ref{NA-Lmdisp}).}
\end{itemize}
To overcome the limitations of the works
\cite{Mamun16,Mamun17,Jannat18a,Jannat18b,Sultana18b,Sultana18c,Chowdhury18,Zaman18,Das19}, we consider a thermally degenerate plasma system [containing thermally degenerate electron species, and non-degenerate warm light and heavy nucleus species, and investigate the arbitrary amplitude DPD NA solitary waves (SWs) by the pseudo-potential approach \cite{Bernstein57,Cairns95}.
The thermally degenerate plasma system under our present consideration is so general that it is valid for hot white dwarfs  \cite{Dufour08,Dufour11,Werner15,Werner19,Koester20} as well as in many space \cite{Rosenberg95,Havnes96,Tsintikidis96,Gelinas98} and laboratory  \cite{Fortov96,Fortov98,Mamun08,Mamun09} plasma environments, where non-degenerate electron-ion plasma with heavy positively charged particles (as impurity or dust) occur.

The structure of the manuscript is as follows. The thermally degenerate plasma model is illustrated in Sec. \ref{model}. The criteria for the existence of subsonic and supersonic DP NA SWs and their basic features for different situations of thermally degenerate plasmas are  investigated by the pseudo-potential approach in Sec. \ref{NA-SWs}. The thermally degenerate plasma model under consideration, results obtained from this investigation, and some important applications are pinpointed as a brief discussion in Sec. \ref{discus}.

\section{\label{model}Model Equations}
We consider a general and realistic TDPS containing the TDES and warm adiabatic degenerate heavy and light nuclei species. We also consider the propagation of thermally degenerate pressure driven (TDPD) nucleus acoustic (NA) waves in such a TDPS. The dynamics of the TDPD NA waves in such a TDPS is described by
\begin{eqnarray}
&&\hspace{-0.4cm}\frac{\partial N_j}{\partial T} +\frac{\partial}{\partial X}(N_jU_j) = 0\,,
\label{be1}\\
&&\hspace{-0.4cm}\frac{\partial\mathcal{P}_{jq}}{\partial T} + U_j\frac{\partial\mathcal{P}_{jq}}{\partial X} +\gamma_j\mathcal{P}_{jq}\frac{\partial U_j}{\partial X} = 0\,,
\label{be2}\\
&&\hspace{-0.4cm}\frac{\partial}{\partial X}(\mathcal{P}_{ed} + \mathcal{P}_{et}) - N_e e\frac{\partial\Phi}{\partial X}=0\,,
\label{be3}\\
&&\hspace{-0.4cm}\frac{\partial U_l}{\partial T}+ U_l\frac{\partial U_l}{\partial X} = -\frac{Z_le}{m_l}\frac{\partial\Phi}{\partial X}-\frac{1}{N_lm_l}\frac{\partial}{\partial X}(\mathcal{P}_{ld}+ \mathcal{P}_{lt})\,,
\label{be4}\\
&&\hspace{-0.4cm}\frac{\partial U_h}{\partial T}+ U_h\frac{\partial U_h}{\partial X} = -\frac{Z_he}{m_h}\frac{\partial\Phi}{\partial X}-\frac{1}{N_hm_h}\frac{\partial}{\partial X}(\mathcal{P}_{hd}+ \mathcal{P}_{ht})\,,
\label{be4-2}\\
&&\hspace{-0.4cm}\frac{\partial^2\Phi}{\partial X^2}=4\pi e(N_e - Z_lN_l - Z_hN_h)\,,
\label{be5}
\end{eqnarray}
where $\Phi$ is the electrostatic NA wave potential; $N_j$ ($U_j$) is number density (fluid speed) of the plasma species $j$ (with $j = e$ for TDES, $j=l$ for degenerate adiabatically warm light nucleus species, and $j=h$ for degenerate adiabatically warm heavy nucleus species;
$\mathcal{P}_{jq}$ in (\ref{be2})$-$(\ref{be4-2}) is the outward pressure for the species $j$ of the type $q$ (with $q=d$ for the degenerate pressure and $q=t$ for the thermal pressure);
$\gamma_j$ is adiabatic index for the plasma species $j$;  $X$ ($T$) is the space (time) variable.

To derive the expression for $\mathcal{P}_{jq}$ from \eqref{be1} and \eqref{be2}, we first make all the dependent
variables to depend only on a single variable $\zeta = X - M T$, where M is the nonlinear wave speed. Now, expressing \eqref{be1} and \eqref{be2} in terms of $\zeta$ and using the steady state condition $\partial/\partial T \rightarrow 0$, we obtain
\begin{eqnarray}
&&-M\frac{dN_j}{d\zeta}+\frac{d}{d\zeta}(N_jU_j) = 0\,,
\label{P1}\\
&&-M\frac{d\mathcal{P}_{jq}}{d\zeta} + U_j\frac{d\mathcal{P}_{jq}}{d\zeta}+ \gamma_j \mathcal{P}_{jq} \frac{dU_j}{d\zeta}= 0\,.
\label{P2}
\end{eqnarray}
Now, integrating \eqref{P1} with respect to $\zeta$ with the appropriate equilibrium conditions (viz. $N_j \rightarrow N_{j0}$ and $U_j \rightarrow 0$), one can write
\begin{equation}\label{P3}
U_j=M\left(1-\frac{N_{j0}}{N_j}\right)\,.
\end{equation}
Inserting \eqref{P3} into \eqref{P2} and dividing the resulting equation by $N_j^{\gamma_j-1}$, we obtain
\begin{equation}\label{P4}
\frac{d}{d\zeta}\left(\frac{{\cal P}_{jq}}{N_j^{\gamma_j}}\right)= 0\,.
\end{equation}
By integrating \eqref{P4} once with respect to $\zeta$, one can express $\mathcal{P}_{jq}$ as
\begin{equation}\label{p}
  \mathcal{P}_{jq} = K_{jq}N_j^{\gamma_j}\,,
\end{equation}
where $K_{jq}=\mathcal{E}_{jq}N_{j0}^{(1-\gamma_j)}$ is the proportionality/integration constant [in which $\mathcal{E}_{jq}$ is equilibrium energy associated with the outward pressure for the species $j$ of type $q$].

We also write the expression for $n_e(=N_e/N_{e0})$ in terms of $\phi(=e\Phi/\mathcal{E}_{e0})$, where
$\mathcal{E}_{e0} = \mathcal{E}_{ed} + \mathcal{E}_{et}$),  as
\begin{equation}
n_e = \left(1+\frac{\gamma_e-1}{\gamma_e}\phi\right)^{\frac{1}{\gamma_e-1}},
\label{ne}
\end{equation}
which  derived by using \eqref{be1}$-$\eqref{be2}. We note that \eqref{ne} is valid for the arbitrary value of $\gamma_e$, and is, thus, valid  for non-relativistically ($\gamma_e=5/3$) as well as ultra-relativistically ($\gamma_e=4/3$) TDES. We also note that for a cold DES, $\mathcal{E}_{et}=0$ and $\mathcal{E}_{e0}=\mathcal{E}_{ed}=K_{ed}N_{e0}^{(\gamma_e-1)}$, which mean that $\phi=e\Phi/\mathcal{E}_{ed}$.  On the other hand, for a non-degenerate thermal electron species,
$\mathcal{E}_{ed}=0$ and $\mathcal{E}_{e0}=\mathcal{E}_{et}=k_BT_e$, which indicate that $\phi=e\Phi/k_BT_e$.

It is worth noting that we cannot directly use $\gamma_e=1$ in \eqref{ne}. To use $\gamma_e=1$ in \eqref{ne}, we expand the latter as
\begin{equation}\label{ne-exp}
n_e=\left(\frac{1}{\gamma_e}\right)\phi + \left(\frac{\gamma_2}{2!\gamma_e^2}\right)\phi^2
             +\left(\frac{\gamma_2\gamma_3}{3!\gamma_e^3}\right)\phi^3+\cdot\cdot\cdot,
\end{equation}
where $\gamma_2 = 2 - \gamma_e$ and $\gamma_3 = 3 - 2\gamma_e$, and by substituting $\gamma_e=1$ into \eqref{ne-exp}, one obtains $n_e$ as
\begin{eqnarray}
&&n_e=1+\phi+\frac{\phi^2}{2!}+\frac{\phi^3}{3!}+\cdot\cdot\cdot=\exp(\phi).
\label{ne-exp}
\end{eqnarray}
Thus, after expressing  (\ref{ne}) in the form of (\ref{ne-exp}), it is valid for $\gamma_e=1$ which yields $n_e=\exp(\phi)$ with $\phi=e\Phi/k_BT_e$.

It is convenient to introduce dimensionless quantities into \eqref{be1}$-$\eqref{be5}. Thus, substituting $P_{ld}$ and $P_{lt}$ as obtained from
\eqref{p} into \eqref{be4} and \eqref{be4-2}, our basic equations \eqref{be1}, \eqref{be4} and \eqref{be4-2} for nucleus species, and the Poisson's equation (\ref{be5}) can be rewritten in dimensionless form as
\begin{eqnarray}
&&\frac{\partial n_l}{\partial t} +\frac{\partial}{\partial x}(n_lu_l) = 0,
\label{be6}\\
&&\frac{\partial n_h}{\partial t} +\frac{\partial}{\partial x}(n_hu_h) = 0,
\label{be6-1}\\
&&\frac{\partial u_l}{\partial t} +u_l\frac{\partial u_l}{\partial x}=-\frac{\partial\phi}{\partial x}-\frac{\sigma_l}{n_l}\frac{\partial n_l^{\gamma_l}}{\partial x},
\label{be7}\\
&&\frac{\partial u_h}{\partial t} +u_h\frac{\partial u_h}{\partial x}=- S_h\frac{\partial\phi}{\partial x}-\frac{\sigma_h}{n_h}\frac{\partial n_h^{\gamma_h}}{\partial x},
\label{be7-1}\\
&&\frac{{\partial}^2\phi}{\partial x^2}=(1+\mu)n_e - n_l - \mu n_h,
\label{be8}
\end{eqnarray}
where we have normalized the variables as $x=X/\lambda_q$, $t=T\omega_{pl}$, $n_l=N_l/N_{l0}$, $n_h=N_h/N_{h0}$, $u_l=U_l/C_q$, $u_h=U_h/C_q$, $\phi=e\Phi/\mathcal{E}_{e0}$, $\sigma_l=\mathcal{E}_{l0}/Z_l\mathcal{E}_{e0}$ (with $\mathcal{E}_{l0}= \mathcal{E}_{ld}+\mathcal{E}_{lt}$), and $\sigma_h=\mathcal{E}_{h0}S_h/Z_h\mathcal{E}_{e0}$. We note that we have redefined $\mathcal{E}_{e0}$, and that the newly defined $\mathcal{E}_{e0}$ must be used in defining
$C_q$ and $\lambda_q$. However, as before $C_q=\lambda_q \omega_{pl}$.

\section{\label{NA-SWs}NA Solitary Waves}
To study arbitrary amplitude TDPD NA SWs, we first assume that all dependent variables in \eqref{be6} -- \eqref{be8} depend on a single independent variable $\xi= x-\mathcal{M}t$, where $\mathcal{M}$ is the the Mach number. This transformation  along with the steady state condition ($\partial /\partial t \rightarrow 0$) leads our basic set of equations to
\begin{eqnarray}
&&\mathcal{M}\frac{dn_l}{d\xi}-\frac{d}{d\xi}(n_lu_l) = 0,
\label{sw1}\\
&&\mathcal{M}\frac{dn_h}{d\xi}-\frac{d}{d\xi}(n_hu_h) = 0,
\label{swh1}\\
&&\mathcal{M}\frac{du_l}{d\xi}-u_l\frac{d u_l}{d\xi}=\frac{d\phi}{d\xi}+\frac{\sigma_l}{n_l}\frac{dn_l^{\gamma_l}}{d\xi},
\label{sw2}\\
&&\mathcal{M}\frac{du_h}{d\xi}-u_h\frac{d u_h}{d\xi}=S_h\frac{d\phi}{d\xi}+\frac{\sigma_h}{n_h}\frac{dn_h^{\gamma_h}}{d\xi},
\label{swh2}\\
&&\frac{d^2\phi}{d\xi^2}=(1+\mu)n_e-n_l-\mu n_h.
\label{sw3}
\end{eqnarray}
Now, by imposing the appropriate boundary conditions (namely, $n_l=1$, $n_h = 1$, $u_l=0$, $u_h = 0$, and $\phi=0$), the integration of  \eqref{sw1}-\eqref{swh2} gives rise to
\begin{eqnarray}
&&u_l=\mathcal{M}\left(1-\frac{1}{n_l}\right),
\label{sw6}\\
&&u_h=\mathcal{M}\left(1-\frac{1}{n_h}\right),
\label{swh6}\\
&&2\mathcal{M}u_l-{u_l}^2-2\phi-\gamma_{\sigma l} [n_l^{(\gamma_l-1)}-1]=0,
\label{sw7}\\
&&2\mathcal{M}u_h-{u_h}^2-2S_h\phi-\gamma_{\sigma h} [n_h^{(\gamma_h-1)}-1]=0,
\label{swh7}
\end{eqnarray}
where $\gamma_{\sigma l} = 2\sigma_l \gamma_l/(\gamma_l-1)$, $\gamma_{\sigma h} = 2\sigma_h \gamma_h/(\gamma_h-1)$ .

Again, substituting $u_l$ and $u_h$ [given by \eqref{sw6} and \eqref{swh6}], respectively, into \eqref{sw7} and \eqref{swh7}, one can obtain equations
for $n_l$ and $n_h$ as
\begin{eqnarray}
&&\gamma_{\sigma l} n_l^{(\gamma_l+1)}-(\mathcal{M}^2+\gamma_{\sigma l} -2\phi)n_l^2 + \mathcal{M}^2=0\,,
\label{sw8}\\
&&\gamma_{\sigma h} n_h^{(\gamma_h+1)}-(\mathcal{M}^2+\gamma_{\sigma h}-2S_h\phi)n_h^2 + \mathcal{M}^2=0\,.
\label{swh8}
\end{eqnarray}
It is important to note that \eqref{sw8} and \eqref{swh8} are valid for the arbitrary value of $\gamma_e$, and $(\gamma_l, \gamma_h) > 1$.  Thus, they can be
used for cold ($\sigma_l = \sigma_h = 0$) as well as adiabatic ($\gamma_l=\gamma_h=3$) non-degenerate light and heavy nucleus species. We also note that we have ignored the the effect of the nucleus degeneracy in our present investigation, because the degeneracy in both light and heavy nuclei species is insignificant compared to that in electron species \cite{Shukla11b,Mamun17-18,Mamun16,Mamun17}.

For the cold light and heavy nucleus species limit ($\sigma_l = \sigma_h = 0$), we can solve (\ref{sw8}) and (\ref{swh8}) for $n_l$ as $n_h$ as
\begin{eqnarray}
&&n_l=\frac{1}{\sqrt{1-\frac{2\phi}{\mathcal{M}^2}}}\,,
\label{nl0}\\
&&n_h=\frac{1}{\sqrt{1-\frac{2S_h\phi}{\mathcal{M}^2}}}.
\label{nh0}
\end{eqnarray}
On the other hand, for both non-degenerate adiabatic light and heavy nucleus species ($\sigma_l=\sigma_{lt}\ne 0$, $\sigma_h=\sigma_{ht}\ne 0$, and $\gamma_l=\gamma_h=3$), \eqref{sw8} and \eqref{swh8} can be expressed, respectively,  as
\begin{eqnarray}
&&3\gamma_l n_l^4-(\mathcal{M}^2+3\sigma_{lt} -2\phi)n_l^2 + \mathcal{M}^2=0,
\label{sw10}\\
&&3\gamma_h n_h^4-(\mathcal{M}^2+3\sigma_{ht} -2S_h\phi)n_h^2 + \mathcal{M}^2=0,
\label{swh0}
\end{eqnarray}
where $\sigma_{lt}=T_{l}/Z_lT_{e}$ and $\sigma_{ht}=S_hT_{h}/Z_hT_{e}$. It is obvious that \eqref{sw10} and \eqref{swh0} are quadratic equations for $n_l^2$ and $n_h^2$, respectively. Therefore, the solution of (\ref{sw10}) and (\ref{swh0}) for $n_l$ and $n_h$ are given by
\begin{eqnarray}
&&n_l=\left[\frac{1}{6\sigma_{lt}}\left(\Phi_{l0}-\sqrt{\Phi_{l0}^2-12\sigma_{lt} \mathcal{M}^2}\right)\right]^{\frac{1}{2}},
 \label{nl1}\\
&&n_h=\left[\frac{1}{6\sigma_{ht}}\left(\Phi_{h0}-\sqrt{\Phi_{h0}^2-12\sigma_{ht} \mathcal{M}^2}\right)\right]^{\frac{1}{2}},
 \label{nh1}
\end{eqnarray}
where $\Phi_{l0}=\mathcal{M}^2+3\sigma_{lt}-2\phi$ and $\Phi_{h0}=\mathcal{M}^2 + 3\sigma_{ht}-2S_h\phi$.

The multiplication of (\ref{sw3}) first by $d\phi/d\xi$, and then the integration of the resulting equation with respect to $\xi$ [under appropriate
boundary conditions, $(d\phi/d\xi)\rightarrow 0$ at $\xi \rightarrow \pm \infty$] give rise to an energy integral in the form
\begin{equation}
\frac{1}{2}\left(\frac{d\phi}{d\xi}\right)^2+V(\phi)=0,
\label{EI}
\end{equation}
where
\begin{equation}
V(\phi)= -\int [(1+\mu)n_e -n_l-\mu n_h]d\phi\,,
\label{PP}
\end{equation}
in which $n_e$ is given by \eqref{ne}. The latter is valid  for $\gamma_e = 5/3$ (non-relativistically TDES) and $\gamma_e=4/3$ (ultra-relativistically TDES), and
\eqref{ne-exp} is valid for $\gamma_e=1$ (Boltzmann distributed electron species). The energy integral (\ref{EI}) [with the pseudo-potential $V(\phi)$ defined by (\ref{PP})] gives rise to the TDPD NA SWs if $[d^2 V/d\phi^2]_{\phi = 0}<0$ so that the fixed point at the origin is unstable \cite{Cairns95} and if at the same time $[d^3 V/d\phi^3]_{\phi = 0}>0~(<0)$ for the TDPD NA SWs with $\phi>0$ ($\phi<0$). We note that $V(0)=0$ and $[dV/d\phi]_{\phi = 0}=0$ are automatically satisfied because of the integration constant chosen and the equilibrium charge neutrality condition, respectively.
We now study the basic features of the TDPD NA SWs for two special situations of TDPS in following two subsections.
\subsection{Cold non-degenerate nucleus species}
We consider here cold non-degenerate nucleus species ($\sigma_l = \sigma_h =0$) which is valid for $(\omega/k)\gg (k_BT_{l0}/m_l)^{1/2}$. Inserting \eqref{ne}, \eqref{nl0}, and \eqref{nh0} into \eqref{PP}, we obtain the pseudo-potential as
\begin{multline}
V(\phi)=C_0-(1+\mu)\left(1+\frac{\gamma_e -1}{\gamma_e}\phi\right)^{\frac{\gamma_e - 1}{\gamma_e}}-\mathcal{M}^2\sqrt{1-\frac{2\phi}{\mathcal{M}^2}}\\
-\frac{\mathcal{M}^2\mu}{S_h}\sqrt{1-\frac{2S_h\phi}{\mathcal{M}^2}},
\label{PP1}
\end{multline}
where $C_0=1 + \mu + \mathcal{M}^2 + \mathcal{M}^2\mu/S_h$ is the integration constant which has been chosen in such a manner that $V(\phi)=0$ at $\phi = 0$.

\begin{figure}[H]
\centering
\includegraphics[width=80mm]{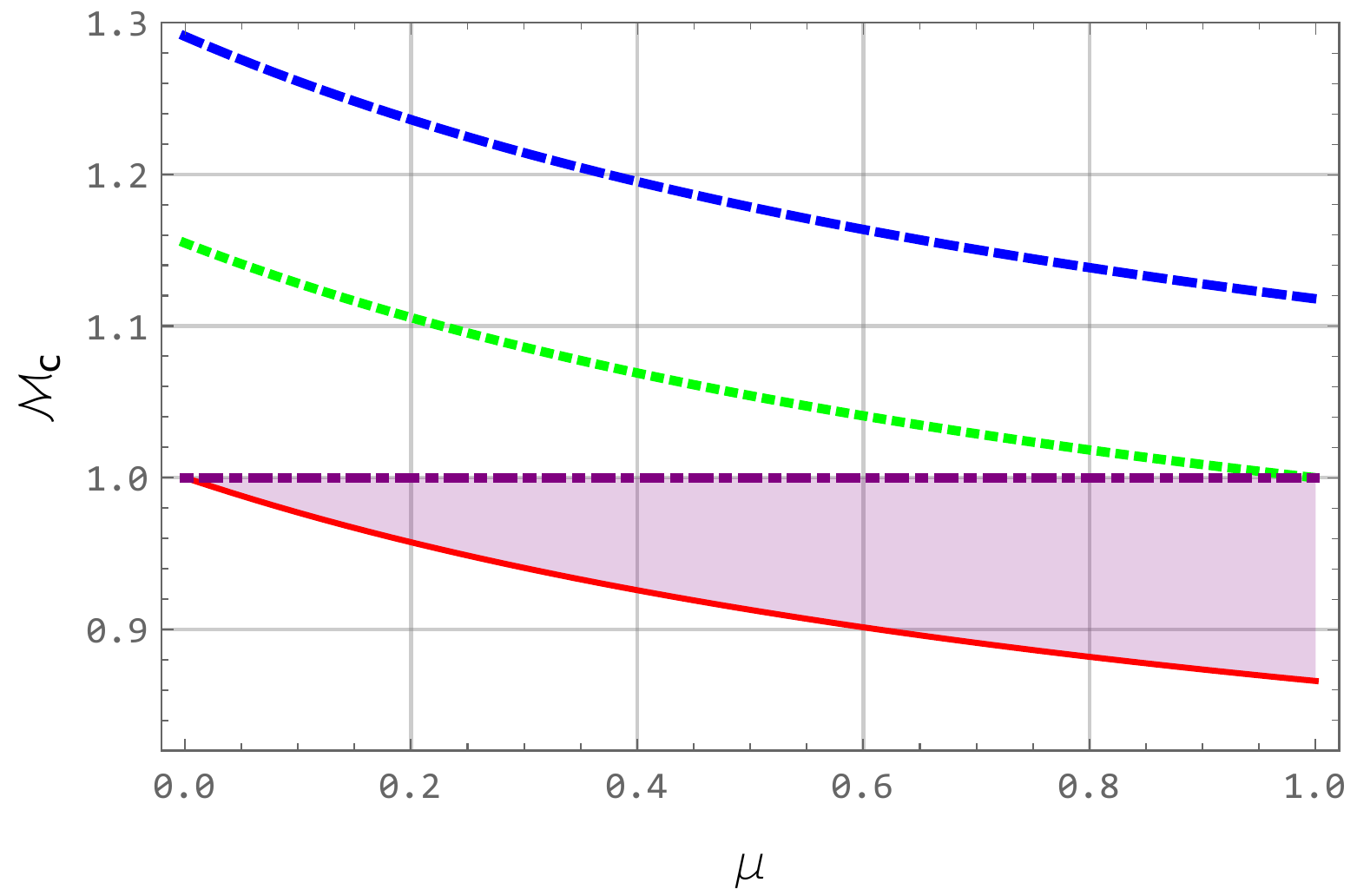}
\caption{The variation of the threshold Mach number $\mathcal{M}_{c}$ with $\mu$ for $S_h = 0.5$, $\gamma_e = 1$ (red solid curve), $\gamma_e = 4/3$ (green dotted curve), and $\gamma_e = 5/3$ (blue dashed curve). The purple dot-dashed line corresponds to $\mathcal{M}_{c} = 1$.}
\label{Fig1}
\end{figure}

To analyze $V(\phi)$ defined by \eqref{PP1}  analytically, for $\phi<0$,  we can expand  $V(\phi)$ as
\begin{equation}
V(\phi) \approx C_2\phi^2+C_3\phi^3+\cdot \cdot \cdot,
\label{PPEXP1}
\end{equation}
where
\begin{eqnarray}
&&C_2=\frac{1}{2!}\left[\frac{1+S_h\mu}{\mathcal{M}^2}-\frac{1}{\gamma_e}(1+\mu)\right],
\label{C2}\\
&&C_3= \frac{1}{3!}\left[\frac{3(1+S_h^2\mu)}{\mathcal{M}^4}-\frac{1}{\gamma_e^2}(2-\gamma_e)(1+\mu)\right].
\label{C3}
\end{eqnarray}
It is obvious from \eqref{PP1} and \eqref{PPEXP1} that $V(\phi) = dV(\phi)/d\phi = 0$ at $\phi = 0$. Therefore, NA solitary wave solution of \eqref{EI} exist if (i) $d^2 V(\phi)/d\phi^2 < 0$ at $\phi = 0$ so that the fixed point at the origin is unstable \cite{Cairns95} and (ii) $[d^3V/d\phi^3]_{\phi=0}>(<)0$ for the NA SWs with $\phi>0$ ($\phi<0$) \cite{Cairns95}. Under the above assumption the NA SWs exist if $C_2<0$, i.e. if $\mathcal{M}>\mathcal{M}_c$, where $\mathcal{M}_c$ is the critical Mach number, which corresponds to the vanishing of the quadratic term in \eqref{PPEXP1}, and is given by
\begin{equation}\label{Mc}
\mathcal{M}_c=\sqrt{\frac{\gamma_e(1+S_h\mu)}{1+\mu}}.
\end{equation}
At this critical value of $\mathcal{M}$, the NA SWs with $\phi>0$ ($\phi<0$) will exist if $C_3 > 0~(<0)$, where $C_3 (\mathcal{M}=\mathcal{M}_c)$ is given by
\begin{equation}
C_3 (\mathcal{M}=\mathcal{M}_c)=\left(\frac{1+\mu}{3!\gamma_e^2}\right)\left[\frac{3(1+\mu)(1+S_h^2\mu)}{(1+S_h\mu)^2}-2+\gamma_e\right].
\end{equation}
It is observed that $C_3(\mathcal{M}=\mathcal{M}_c)>0$ for $\mu\ge0$, $S_h > 0$ and $\gamma_e\ge 1$. Therefore, our plasma system under consideration only supports the NA SWs with $\phi>0$ for any possible values of $\mu$, $S_h$, and $\gamma_e$. Figure \ref{Fig1} shows how the critical Mach number $\mathcal{M}_c$ varies with $\mu$ for the isothermal electron species $\gamma_e=1$ (red solid curve), ultra-relativistically DES $\gamma_e=4/3$ (green dotted curve), and non-relativistically DES $\gamma_e=5/3$ (blue dashed curve). It is seen that as the non-degenerate heavy nucleus number density increases, the critical Mach number ($\mathcal{M}_c$) decreases. It also indicates that the isothermal electron species supports the formation of both subsonic and supersonic NA SWs. The existence of subsonic NA SWs region is represented by the shadow area, as shown in Fig. \ref{Fig1}. This region becomes broader with the increase in $\mu$. The supersonic NA SWs region is found above the purple dot-dashed line ($\mathcal{M}_c =1$). On the other hand, the ultra-relativistic and non-relativistic DES support only the supersonic NA SWs for $0<\mu<1$.

\begin{figure}[H]
\centering
\includegraphics[width=80mm]{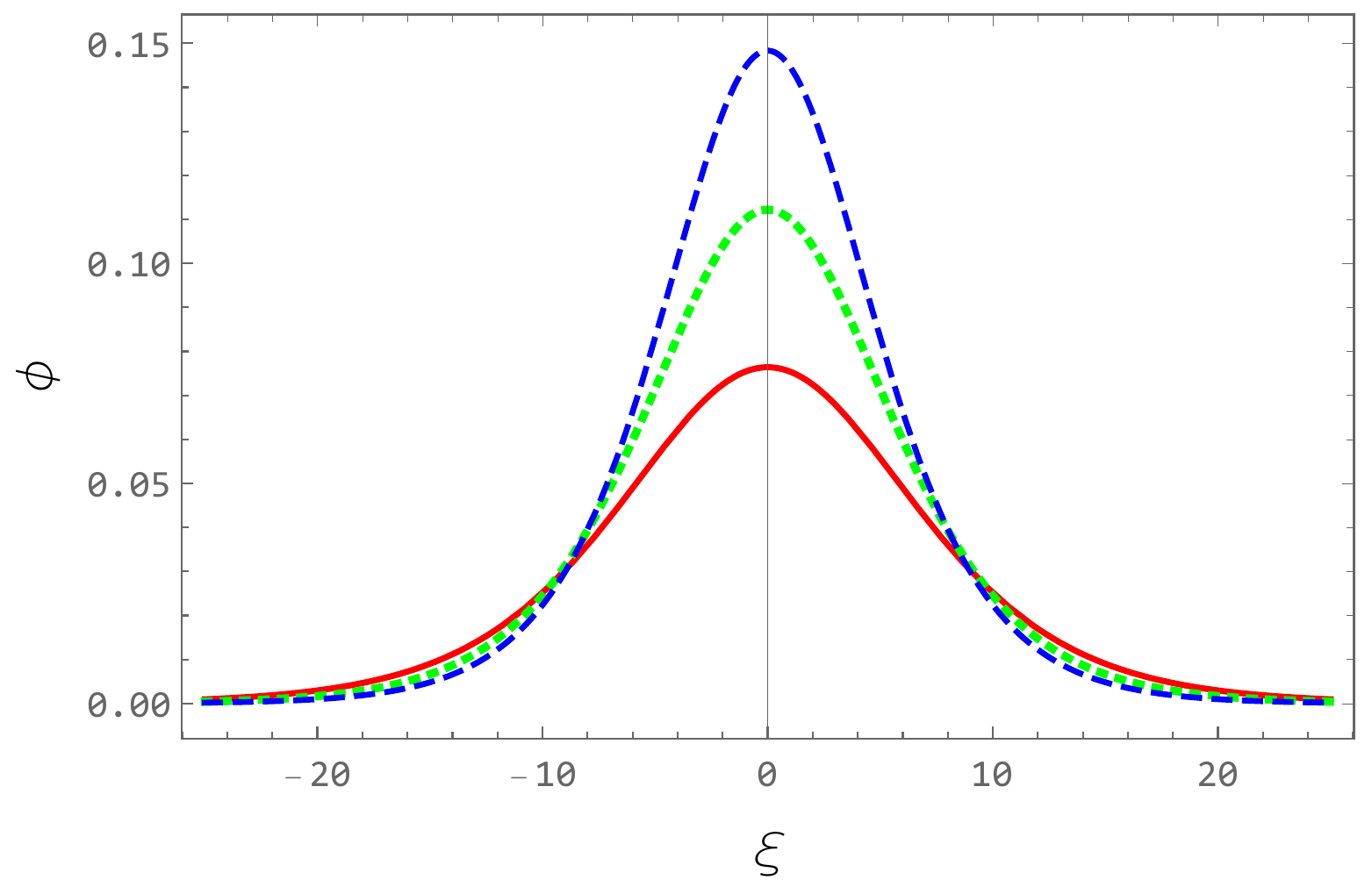}
\caption{The variation of the small amplitude subsonic NA SWs for different values of $\mu = 0.15$ (red solid curve), $\mu = 0.2$ (green dotted curve), and $\mu = 0.25$ (blue dashed curve) at $\gamma_e = 1$, $\mathcal{M} = 0.99$, and $S_h = 0.5$.}
\label{Fig2}
\end{figure}

\begin{figure}[H]
\centering
\includegraphics[width=80mm]{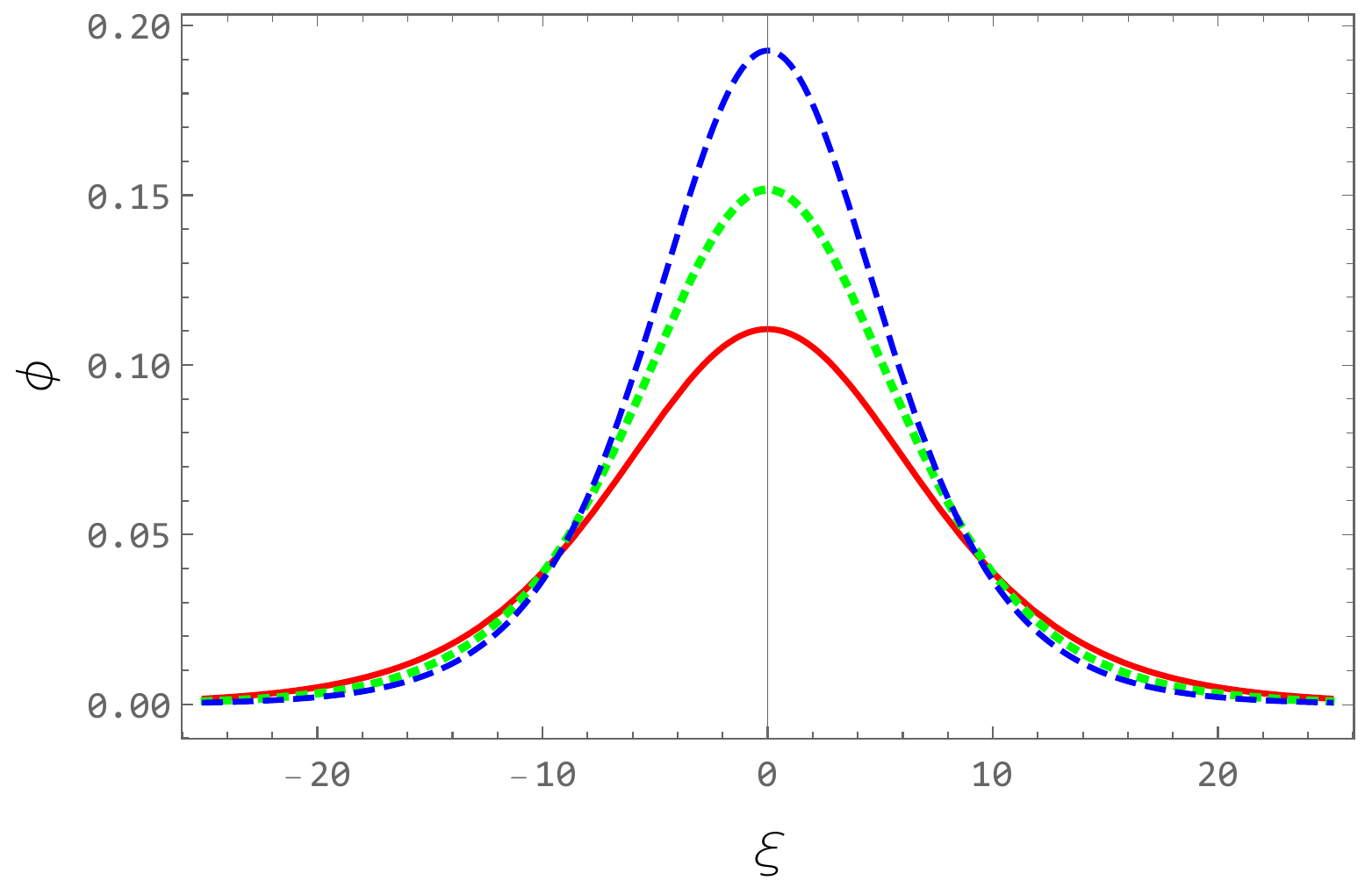}
\caption{The variation of the small amplitude supersonic NA SWs for different values of $\mu = 0.15$ (red solid curve), $\mu = 0.2$ (green dotted curve), and $\mu = 0.25$ (blue dashed curve) at $\gamma_e = 4/3$, $\mathcal{M} = 1.15$, and $S_h = 0.5$.}
\label{Fig3}
\end{figure}

\begin{figure}[H]
\centering
\includegraphics[width=80mm]{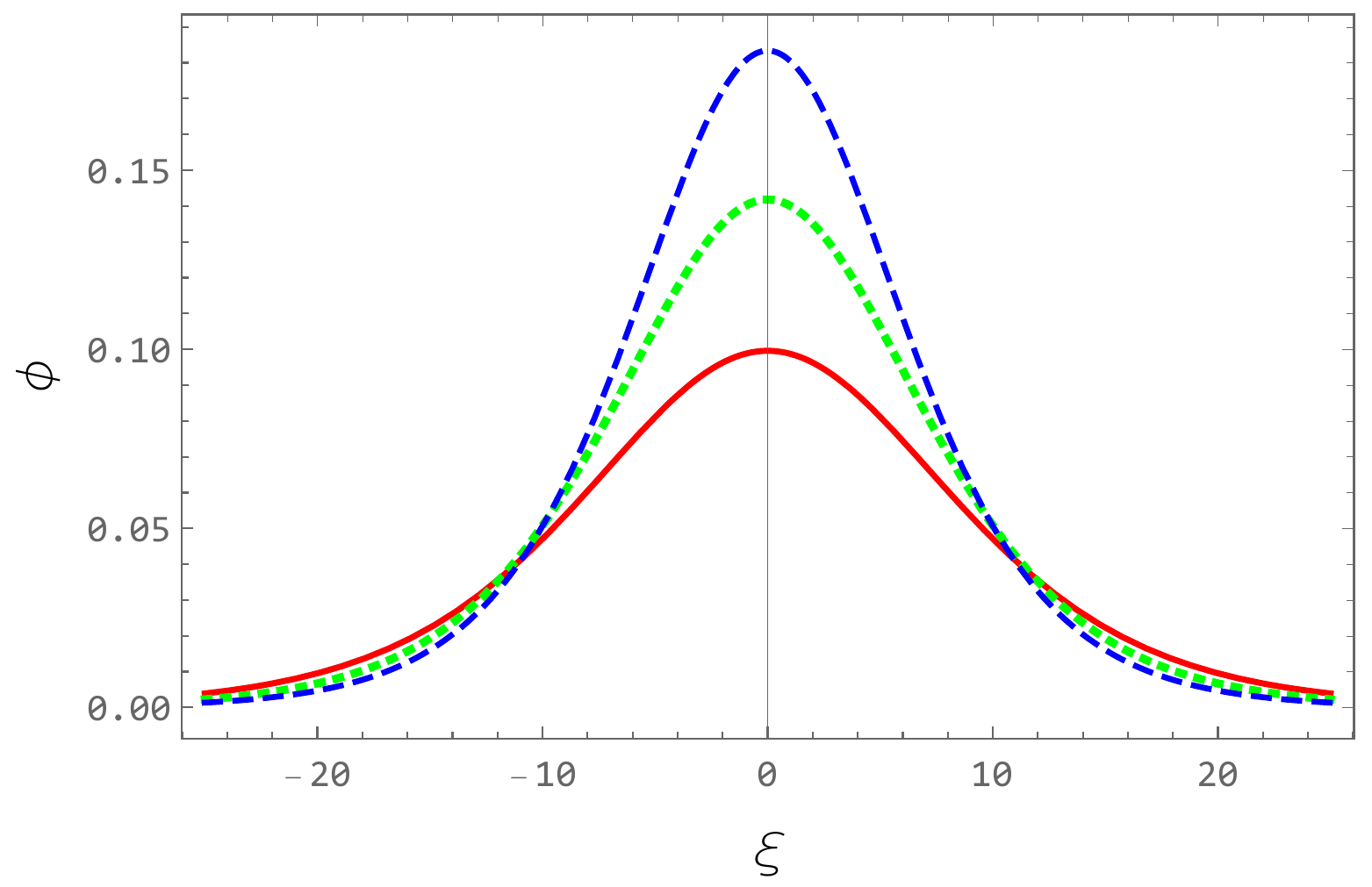}
\caption{The variation of the small amplitude supersonic NA SWs for different values of $\mu = 0.15$ (red solid curve), $\mu = 0.2$ (green dotted curve), and $\mu = 0.25$ (blue dashed curve) at $\gamma_e = 5/3$, $\mathcal{M} = 1.28$, and $S_h = 0.5$.}
\label{Fig4}
\end{figure}

\begin{figure}[H]
\centering
\includegraphics[width=80mm]{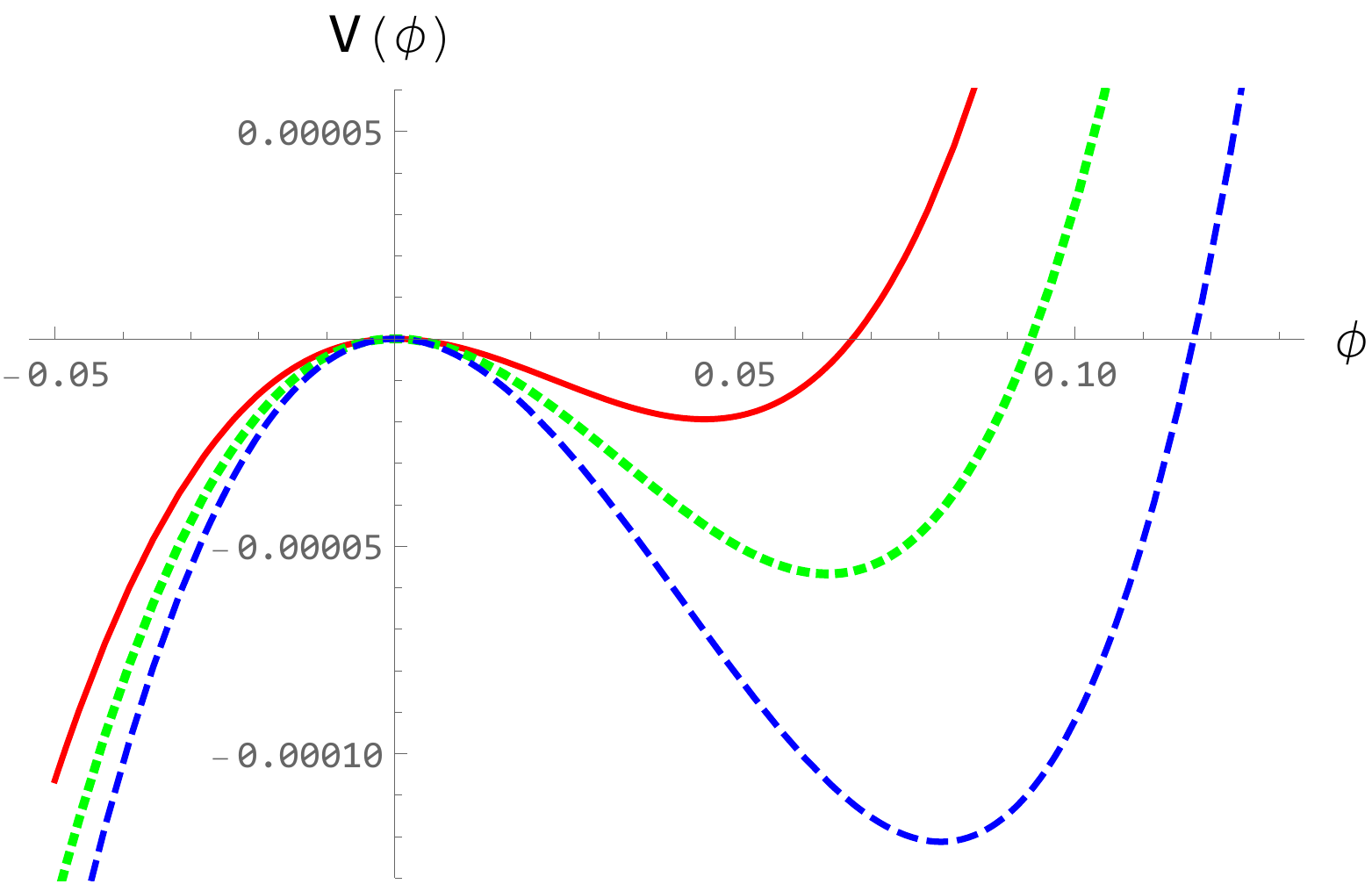}
\caption{The formation of potential wells in positive $\phi$-axis for $\mu = 0.15$ (red solid curve), $\mu = 0.2$ (green dotted curve), and $\mu = 0.25$ (blue dashed curve) at $\gamma_e = 1$, $\mathcal{M} = 0.99$, and $S_h = 0.5$.}
\label{Fig5}
\end{figure}

\begin{figure}[H]
\centering
\includegraphics[width=80mm]{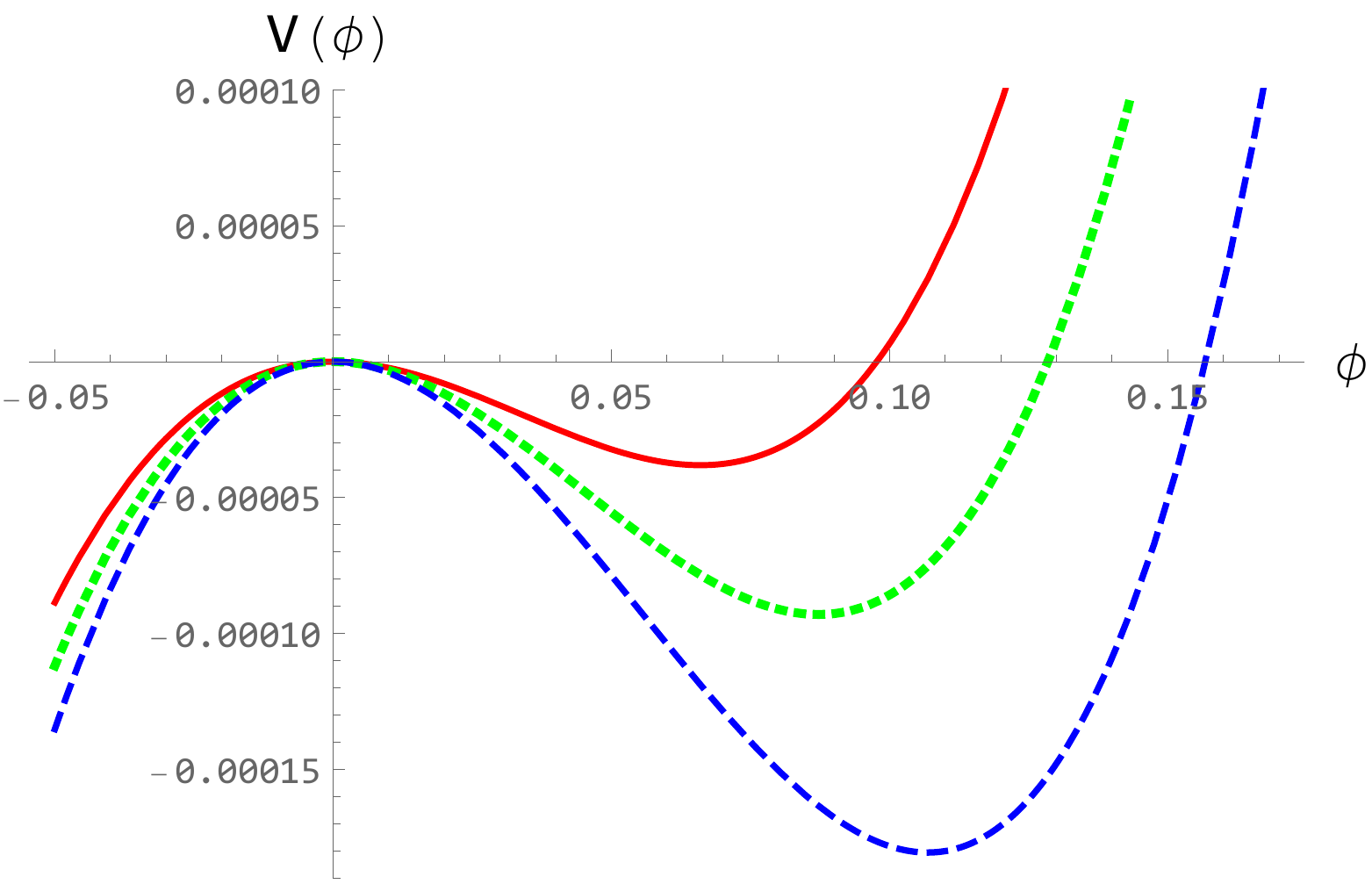}
\caption{The formation of potential wells in positive $\phi$-axis for $\mu = 0.15$ (red solid curve), $\mu = 0.2$ (green dotted curve), and $\mu = 0.25$ (blue dashed curve) at $\gamma_e = 4/3$, $\mathcal{M} = 1.15$, and $S_h = 0.5$.}
\label{Fig6}
\end{figure}

\begin{figure}[H]
\centering
\includegraphics[width=82mm]{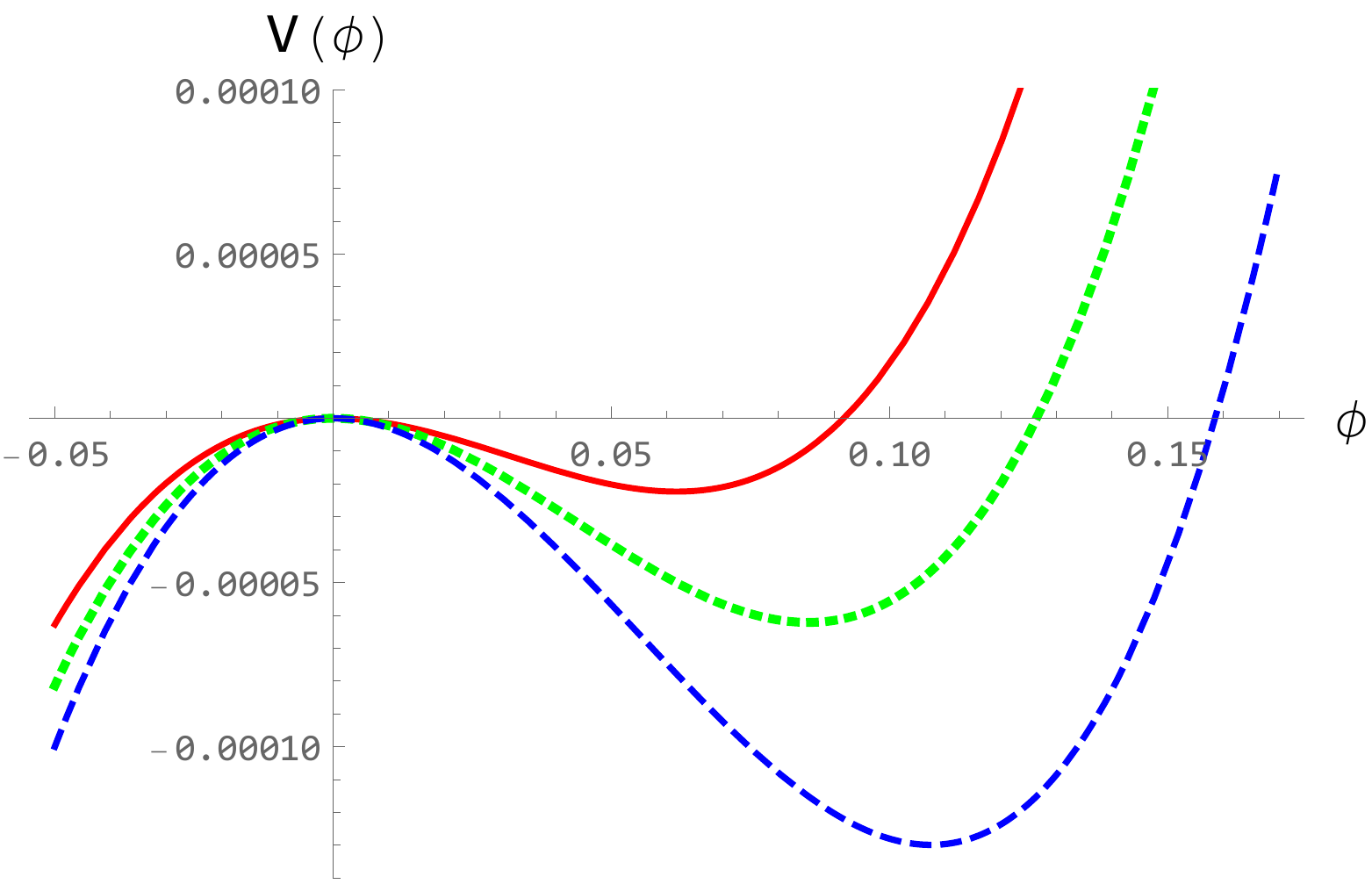}
\caption{The formation of potential wells in positive $\phi$-axis for $\mu = 0.15$ (red solid curve), $\mu = 0.2$ (green dotted curve), and $\mu = 0.25$ (blue dashed curve) at $\gamma_e = 5/3$, $\mathcal{M} = 1.28$, and $S_h = 0.5$.}
\label{Fig7}
\end{figure}

We first investigate the properties of small amplitude NA SWs by considering the approximation [given by \eqref{PPEXP1}]. Inserting \eqref{PPEXP1} into \eqref{EI} and upon integrating along with the condition $V(\phi) = 0$ at $\phi \rightarrow \phi_{m}$, we obtain, in the small amplitude limit, the NA solitary wave solution
\cite{Mamun08}
\begin{equation}\label{sol}
  \phi = \left(-\frac{C_2}{C_3}\right)\text{sech}^2\left(\sqrt{-\frac{C_2}{2}}\xi\right)\,.
\end{equation}
The profiles (indicating the amplitude and width) of the small amplitude subsonic ($\mathcal{M}_c < \mathcal{M} < 1$) and supersonic ($\mathcal{M} >1$ and $\mathcal{M}>\mathcal{M}_c$) NA SWs associated with the positive potential are graphically displayed in Figs. \ref{Fig2} - \ref{Fig4}. We also investigate the properties of arbitrary amplitude NA SWs by numerical analyses of \eqref{PP1}. Our direct numerical analysis of \eqref{PP1} also show the existence of positive NA SWs potential. Figures \ref{Fig5} - \ref{Fig7} displays the formation of the potential wells in the positive $\phi$-axis for the same set of plasma parameters as that in small amplitude limit. It is found for the small amplitude limit that the subsonic NA SWs with $\phi>0$ exist for the non-degenerate isothermal electron, but both the ultra-relativistic and non-relativistic degenerate electron supports the supersonic NA SWs with $\phi>0$. It is observed that the amplitude (width) of the NA SWs increases (decreases) as the number density of heavy nucleus species increases. Thus, the effect of the ultra-relativistic degenerate electron significantly modifies the basic features of NA SWs. It is found that the amplitude of NA SWs in the non-relativistically DES is much smaller than that in ultra-relativistically DES, but is larger than that in Boltzmann distributed electron species (BDES). Note that the width of supersonic NA SWs in ultra-relativistically degenerate electron species is much wider than that in both other electron species. On the other hand, for arbitrary amplitude limit Figs. \ref{Fig5} - \ref{Fig7} provide a visualization of the amplitude ($\phi_m$), which is the intercept on the positive $\phi$-axis, and the width ($\phi_m/\sqrt{|V_m|}$, where
$|V_m|$ is the maximum value of $V(\phi)$ in the potential wells formed in the positive $\phi$-axis.
The increase in $\mu$ causes to increase (decrease) the amplitude (width) of both subsonic and supersonic NA SWs. The depth of potential wells for the ultra-relativistically DES is much larger than that in isothermal and non-relativistically electron species. The effects of $\gamma_e$ shows the similar results as that in the case of small amplitude limit. It is concluded from this visualization that the variation of the amplitude and the width with $\mu$ in the case of arbitrary amplitude NA SWs is almost the same as that in the case of small amplitude NA SWs.
\subsection{Adiabatically warm non-degenerate nucleus species}
\begin{figure}[H]
\centering
\includegraphics[width=80mm]{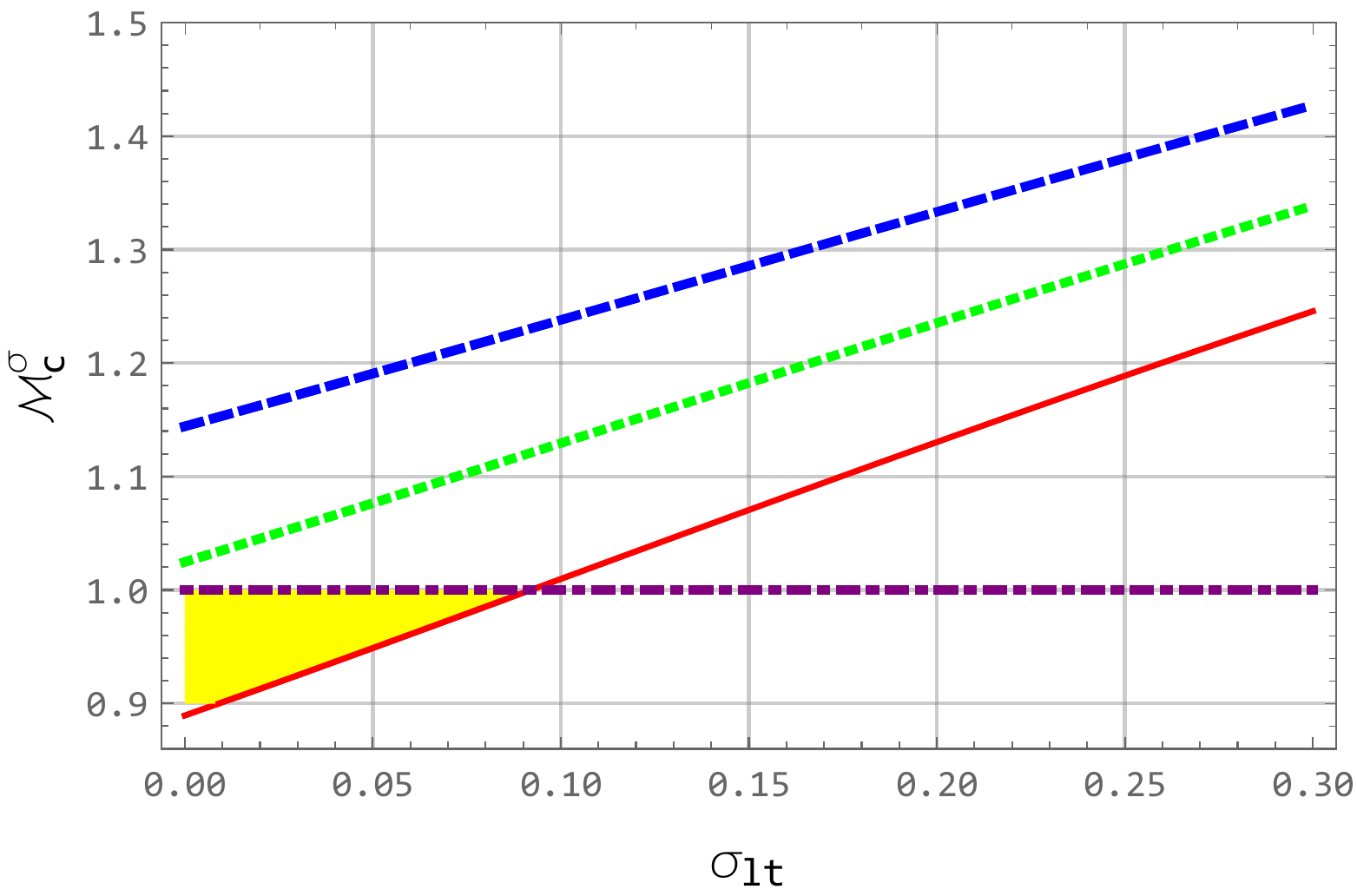}
\caption{The variation of the threshold Mach number $\mathcal{M}_{c}^{\sigma}$ with $\sigma_{lt}$ for $\mu = 0.8$, $S_h = 0.5$, $\sigma_{ht} = 0.015$, $\gamma_e = 1$ (red solid curve), $\gamma_e = 4/3$ (green dotted curve), and $\gamma_e = 5/3$ (blue dashed curve). The purple dot-dashed line corresponds to $\mathcal{M}_{c}^{\sigma} = 1$.}
\label{Fig8}
\end{figure}

We finally consider non-degenerate warm adiabatic nucleus species where the light [heavy] number density defined by \eqref{nl1} [\eqref{nh1}]. The nucleus number densities [given by \eqref{nl1} and \eqref{nh1}] are valid when $P_{jd} \ll P_{jt}$ which is valid not only for hot white dwarfs \cite{Dufour08,Dufour11,Werner15,Werner19,Koester20}, but also for many space \cite{Rosenberg95,Havnes96,Tsintikidis96,Gelinas98} and laboratory \cite{Fortov96,Fortov98} plasma environments. Now, inserting \eqref{ne}, \eqref{nl1}, and \eqref{nh1} into \eqref{PP}, and following the same procedure as mentioned before, we can obtain the pseudo-potential $V(\phi)$ as
\begin{eqnarray}
V(\phi)=C_0^\sigma-(1+\mu)\left[1+\left(\frac{\gamma_e-1}{\gamma_e}\right)\phi\right]^{\frac{\gamma_e}{\gamma_e-1}} \nonumber\\
-\frac{\sqrt{2}}{3\sqrt{3\sigma_{lt}}}\left(\sqrt{\Phi_{l0}-\Phi_{l1}}\right)\left(\Phi_{l0}+\frac{1}{2} \Phi_{l1}\right)\nonumber\\
-\frac{\mu\sqrt{2}}{3S_h\sqrt{3\sigma_{ht}}}\left(\sqrt{\Phi_{h0}-\Phi_{h1}}\right)\left(\Phi_{h0}+\frac{1}{2} \Phi_{h1}\right),
\label{PP2}
\end{eqnarray}
where  $C_0^\sigma=1 +\mu +\sigma_{lt}+\mathcal{M}^2 + \mu(\mathcal{M}^2 + \sigma_{ht})/S_h$ is the integration constant chosen in such a way that $V(\phi)=0$ at $\phi=0$, $\Phi_{l0}=\mathcal{M}^2+3\sigma_{lt}-2\phi$, $\Phi_{l1}=\sqrt{\Phi_{l0}^2-12\sigma_{lt} \mathcal{M}^2}$, $\Phi_{h0}=\mathcal{M}^2+3\sigma_{ht}-2S_h\phi$, and $\Phi_{h1}=\sqrt{\Phi_{h0}^2-12\sigma_{ht} \mathcal{M}^2}$. To find the solitary wave solution of \eqref{EI}, the pseudo-potential $V(\phi)$ must satisfy the necessary conditions as mentioned before. Therefore, to find the conditions for the existence of the NA SWs, we expand $V(\phi)$ as
\begin{eqnarray}
V(\phi)\approx C_2^\sigma\phi^2+ C_3^\sigma\phi^3 + \cdot \cdot \cdot,
\label{PPEXP2}
\end{eqnarray}
where
\begin{eqnarray}
&&C_2^\sigma=\frac{1}{2!}\left[\frac{S_h\mu}{\mathcal{M}^2-3\sigma_{ht}}+\frac{1}{\mathcal{M}^2-3\sigma_{lt}} -\frac{1}{\gamma_e}(1+\mu)\right],
\label{C2s}\\
&&C_3^\sigma=\frac{1}{3!}\left[\frac{3S_h^2\mu(\mathcal{M}^2+\sigma_{ht})}{(\mathcal{M}^2-3\sigma_{ht})^3}+\frac{3(\mathcal{M}^2+\sigma_{lt})}{(\mathcal{M}^2-3\sigma_{lt})^3}\right.\nonumber\\
&&\hspace{2mm}\left.-\frac{1}{\gamma_e^2}(2-\gamma_e)(1+\mu)\right].
\label{C3s}
\end{eqnarray}
The coefficient of $\phi^2$ (viz. $C_2^\sigma$) indicates from $[d^2V/d\phi^2]_{\phi=0}<0$) that the solitary wave solution of (\ref{EI}) with (\ref{PP2}) exists if and only if $C_2^\sigma<0$. Thus, the NA SWs exist if $\mathcal{M}>\mathcal{M}_c^\sigma$, where
$\mathcal{M}_c^\sigma$ is given by
\begin{equation}
\mathcal{M}_c^\sigma=\left(\frac{b+\sqrt{b^2-4ac}}{2a}\right)^{1/2},
\label{Mcs}
\end{equation}
$a = 1+\mu$, $b = 3a(\sigma_{lt} + \sigma_{ht}) + \gamma_e(S_h\mu + 1)$, and $c = 3\gamma_e(S_h\mu\sigma_{lt}+\sigma_{ht}) + 9a\sigma_{lt}\sigma_{ht}$. We get $\mathcal{M}_c^\sigma=\mathcal{M}_c$ if we neglect the temperature of light and heavy ions species (i.e. $\sigma_{lt} = \sigma_{ht} = 0$). On the other hand, the NA SWs exist with $\phi>0$ ($\phi<0$) if $C_3 (\mathcal{M}=\mathcal{M}_c^\sigma)>0~(<0)$. It has been checked that $C_3 (\mathcal{M}=\mathcal{M}_c^\sigma)>0$ for $\mu\ge 0$, $\sigma_{lt}\ge 0$, $\sigma_{ht}\ge 0$, and $\gamma_e\ge 1$. Therefore, the NA SWs only with $\phi>0$ exist for all possible values of $\mu$, $\sigma_{lt}$, $\sigma_{ht}$, and $\gamma_e$. Figure \ref{Fig8} displays how the critical Mach number ($\mathcal{M}_c^\sigma$) varies with $\sigma_{lt}$ for $\mu=0.8$. It is seen that $\mathcal{M}_c^\sigma$ increases with $\sigma_{lt}$ for all the possible values of $\gamma_e$. The effects of the temperature of light and heavy nucleus species reduce the region where the subsonic NA SWs exist. In the presence of warm adiabatic light and heavy nuclei species, the region of subsonic SWs shrinks as the number density of light nucleus species increases. The similar effect of $\mathcal{M}_c^\sigma$ with $\sigma_{ht}$ has been observed (which is not shown here). The non-relativistically and ultra-relativistically degenerate electrons as well as the temperature of light and heavy nuclei species are also here against the formation of subsonic NA SWs, but are in favor of the formation of supersonic NA SWs with $\phi>0$.

For the small amplitude limit, the solitary wave solution of \eqref{EI} with the approximation [given by \eqref{PPEXP2}] as well as the condition $V(\phi) = 0$ at $\phi \rightarrow \phi_{m}$ can be written as \cite{Mamun08}
\begin{equation}\label{sol2}
  \phi = \left(-\frac{C_2^{\sigma}}{C_3^{\sigma}}\right)\text{sech}^2\left(\sqrt{-\frac{C_2^{\sigma}}{2}}\xi\right)\,.
\end{equation}
To study the role of nucleus temperature ($\sigma_{lt}, \,\sigma_{ht}$) on the basic properties of both large and small amplitudes subsonic and supersonic NA solitary structures, we visualize the solution \eqref{sol2} and numerically solve the pseudo-potential $V(\phi)$ [given by \eqref{PP2}] for $\gamma_e = 1$ (BDES), $\gamma_e = 4/3$ (ultra-relativistically DES), and $\gamma_e = 5/3$ (non-relativistically DES). 
\begin{figure}[H]
\centering
\includegraphics[width=78mm]{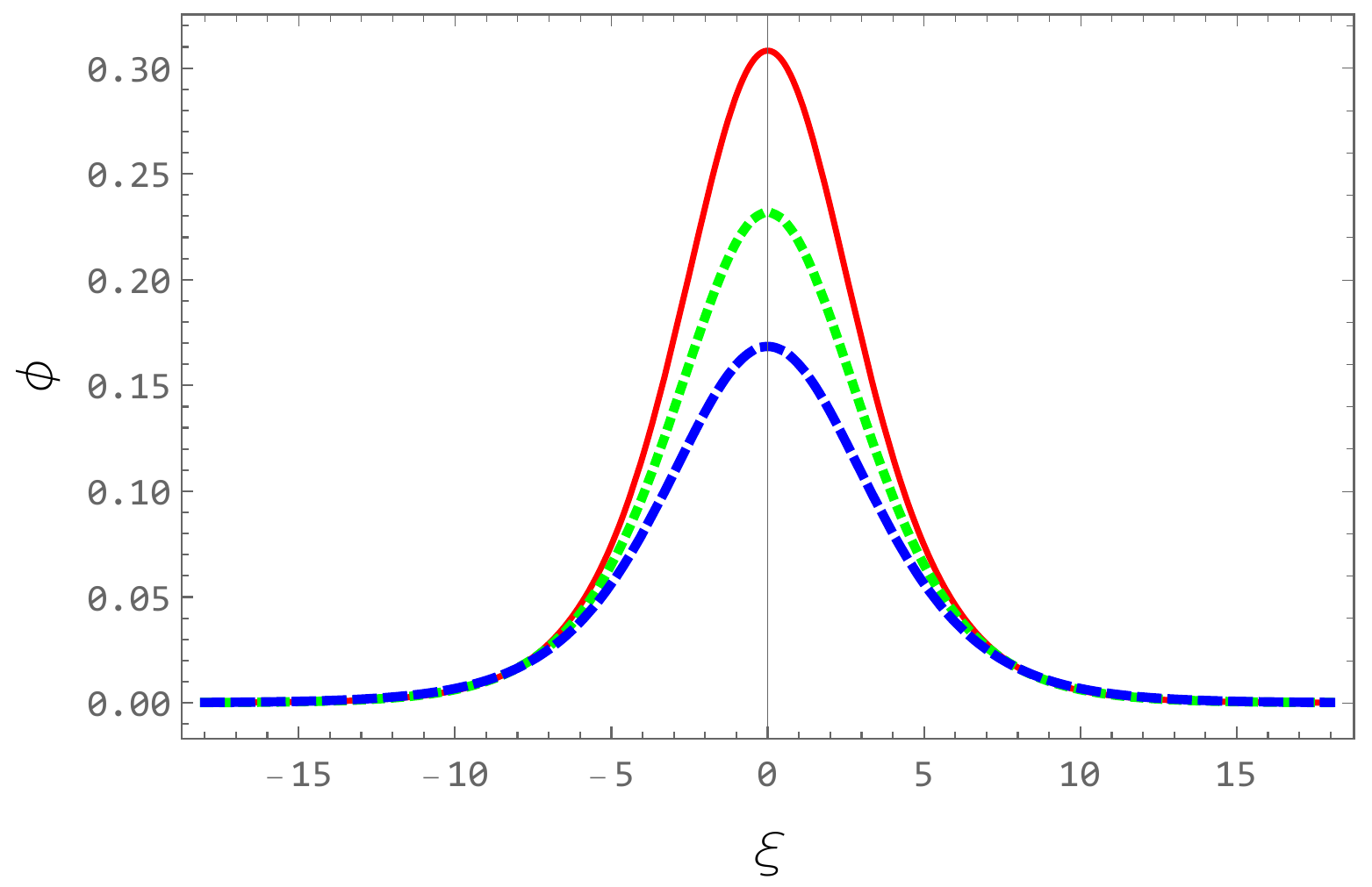}
\caption{The variation of the small amplitude subsonic NA SWs for different values of $\sigma_{lt} = 0.02$ (red solid curve), $\sigma_{lt} = 0.03$ (green dotted curve), and $\sigma_{lt} = 0.04$ (blue dashed curve) at $\gamma_e = 1$, $\mathcal{M} = 0.99$, $\mu = 0.8$, $S_h = 0.5$, and $\sigma_{ht} = 0.015$.}
\label{Fig9}
\end{figure}
\begin{figure}[H]
\centering
\includegraphics[width=80mm]{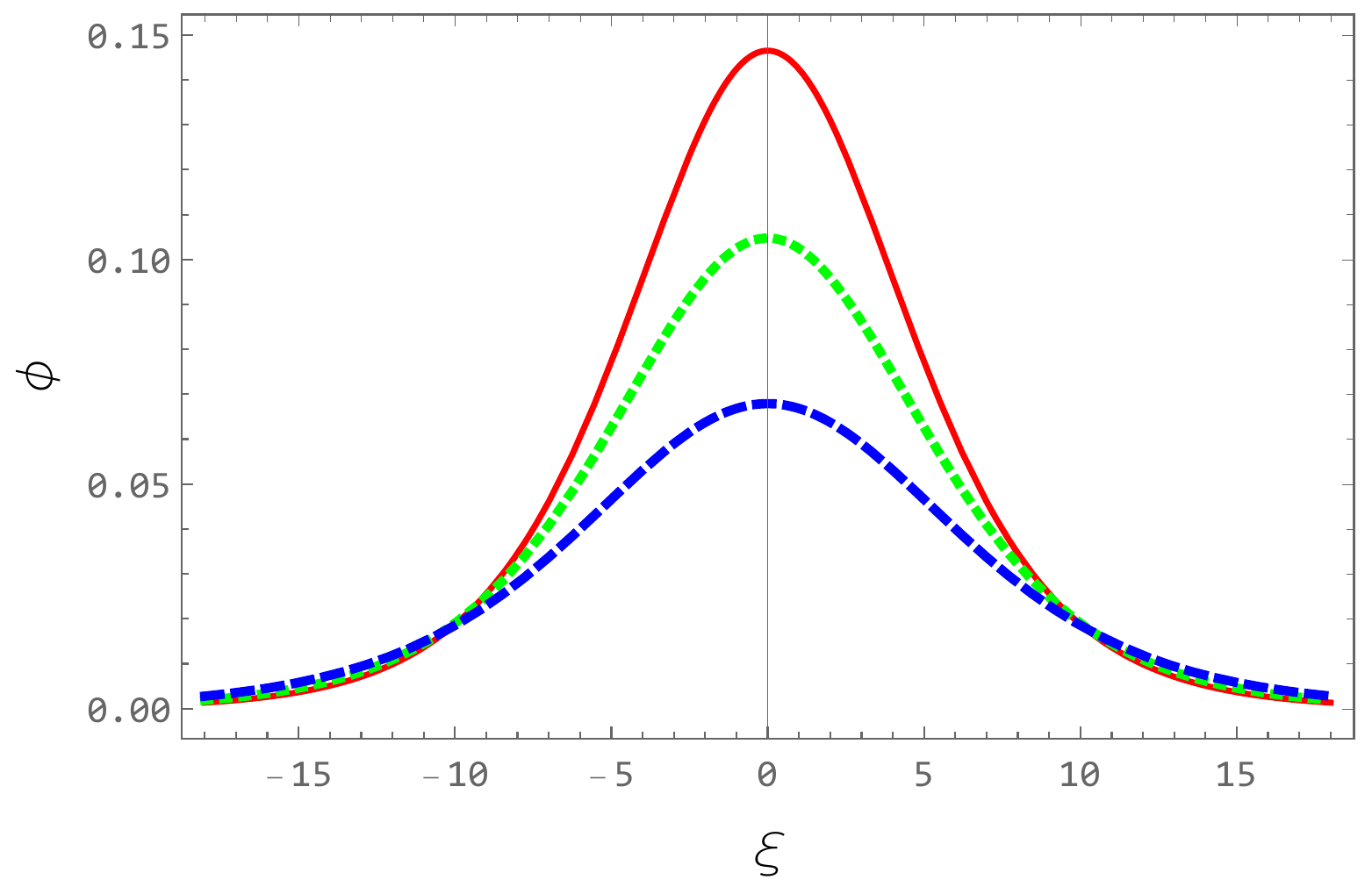}
\caption{The variation of the small amplitude supersonic NA SWs for different values of $\sigma_{lt} = 0.02$ (red solid curve), $\sigma_{lt} = 0.03$ (green dotted curve), and $\sigma_{lt} = 0.04$ (blue dashed curve) at $\gamma_e = 4/3$, $\mathcal{M} = 1.09$, $\mu = 0.8$, $S_h = 0.5$, and $\sigma_{ht} = 0.015$.}
\label{Fig10}
\end{figure}
\begin{figure}[H]
\centering
\includegraphics[width=80mm]{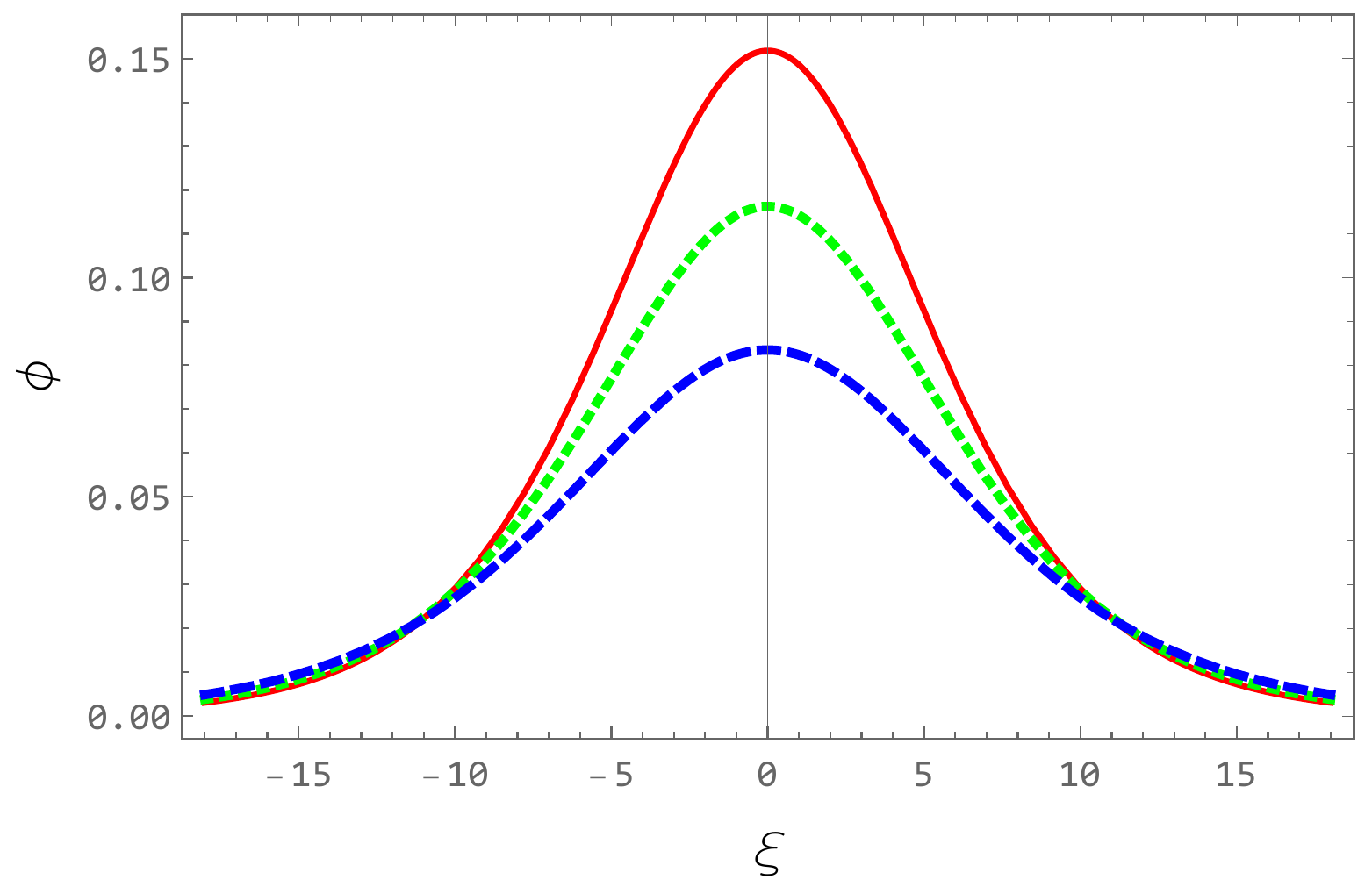}
\caption{The variation of the small amplitude supersonic NA SWs for different values of $\sigma_{lt} = 0.02$ (red solid curve), $\sigma_{lt} = 0.03$ (green dotted curve), and $\sigma_{lt} = 0.04$ (blue dashed curve) at $\gamma_e = 5/3$, $\mathcal{M} = 1.21$, $\mu = 0.8$, $S_h = 0.5$, and $\sigma_{ht} = 0.015$.}
\label{Fig11}
\end{figure}
\begin{figure}[H]
\centering
\includegraphics[width=80mm]{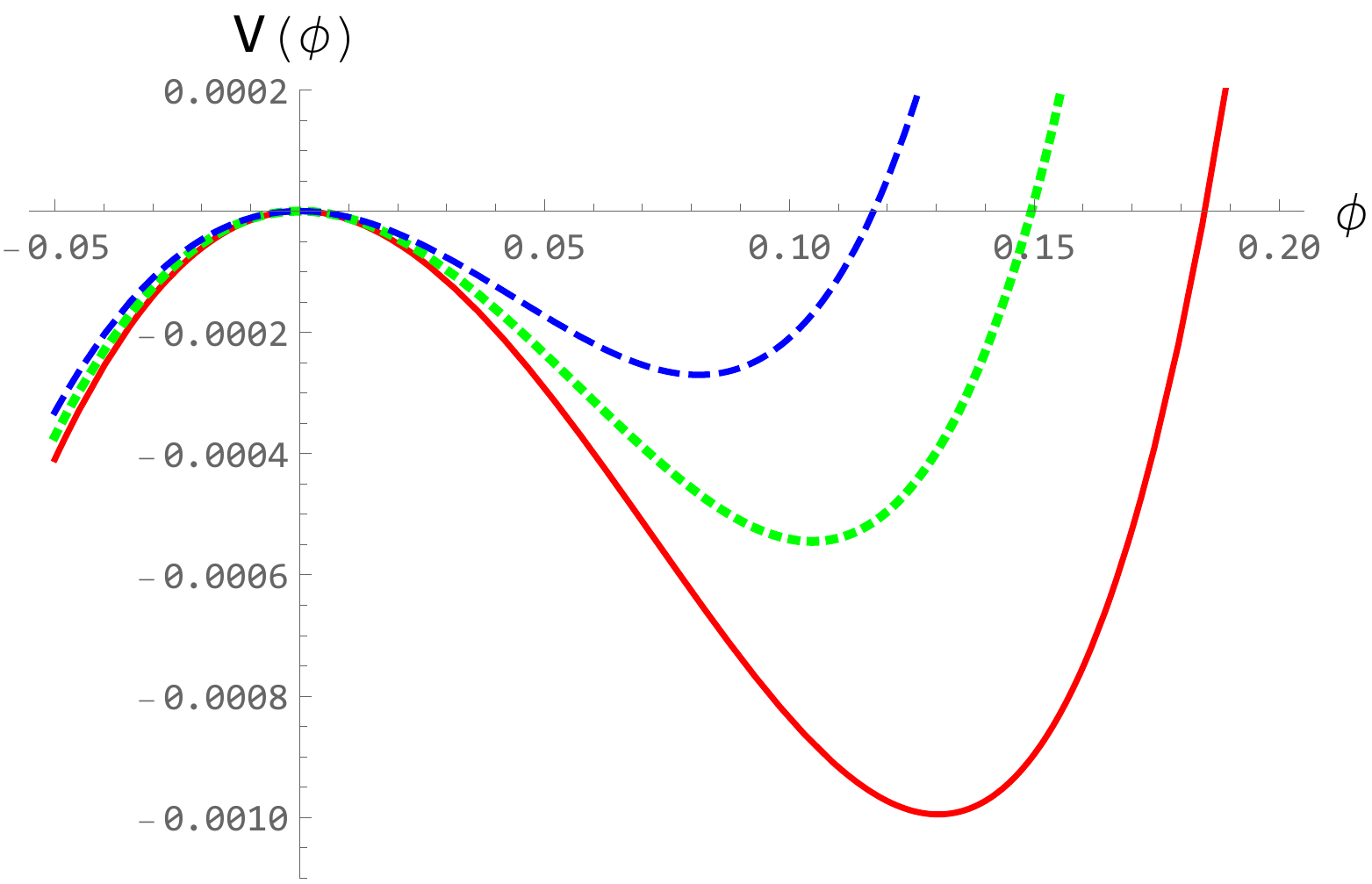}
\caption{The formation of potential wells in positive $\phi$-axis for $\sigma_{lt} = 0.02$ (red solid curve), $\sigma_{lt} = 0.03$ (green dotted curve), and $\sigma_{lt} = 0.04$ (blue dashed curve) at $\gamma_e = 1$, $\mathcal{M} = 0.99$, $\mu = 0.8$, $S_h = 0.5$, and $\sigma_{ht} = 0.015$.}
\label{Fig12}
\end{figure}
\begin{figure}[H]
\centering
\includegraphics[width=80mm]{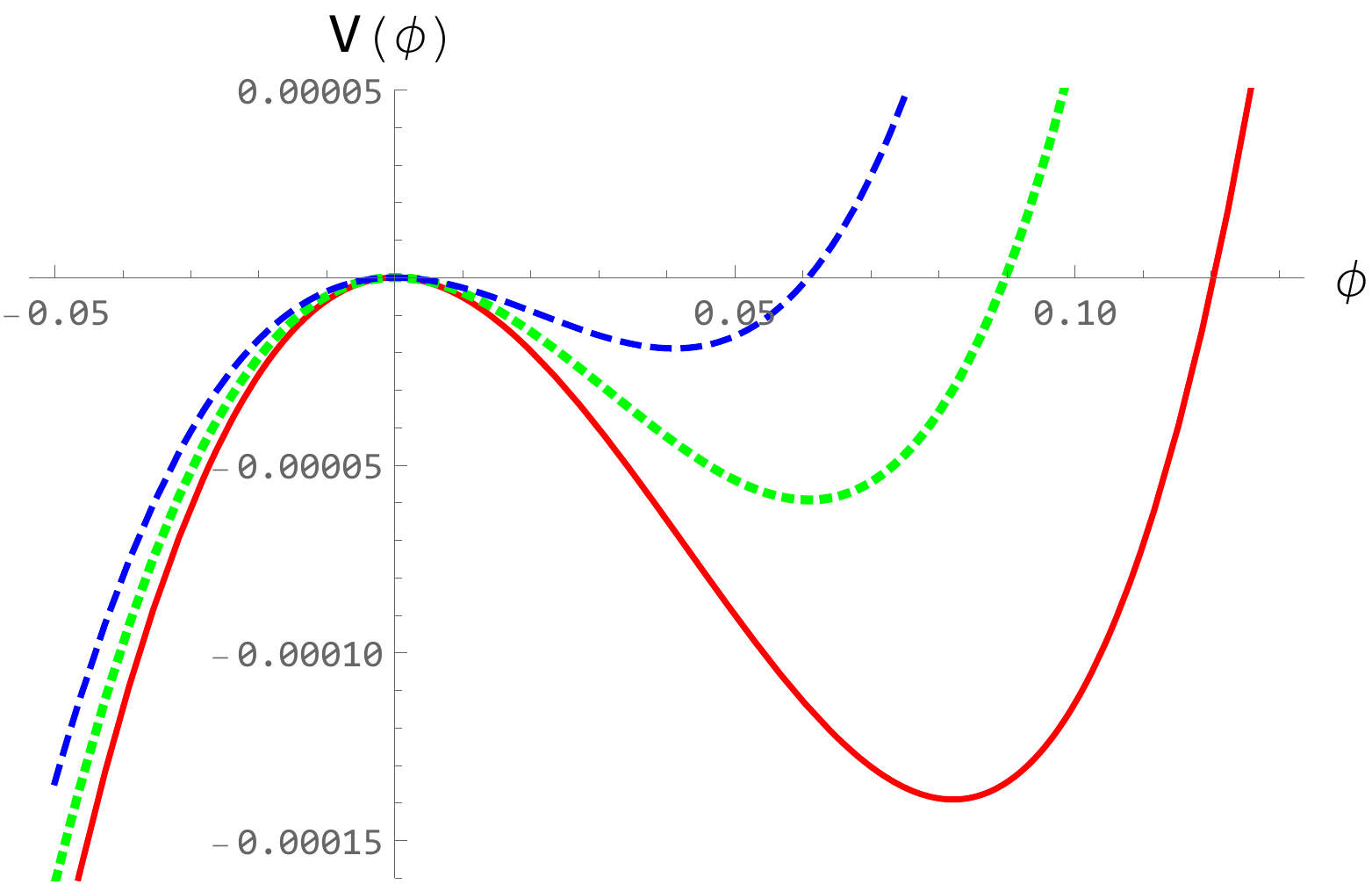}
\caption{The formation of potential wells in positive $\phi$-axis for $\sigma_{lt} = 0.02$ (red solid curve), $\sigma_{lt} = 0.03$ (green dotted curve), and $\sigma_{lt} = 0.04$ (blue dashed curve) at $\gamma_e = 4/3$, $\mathcal{M} = 1.09$, $\mu = 0.8$, $S_h = 0.5$, and $\sigma_{ht} = 0.015$. }
\label{Fig13}
\end{figure}
\begin{figure}[H]
\centering
\includegraphics[width=80mm]{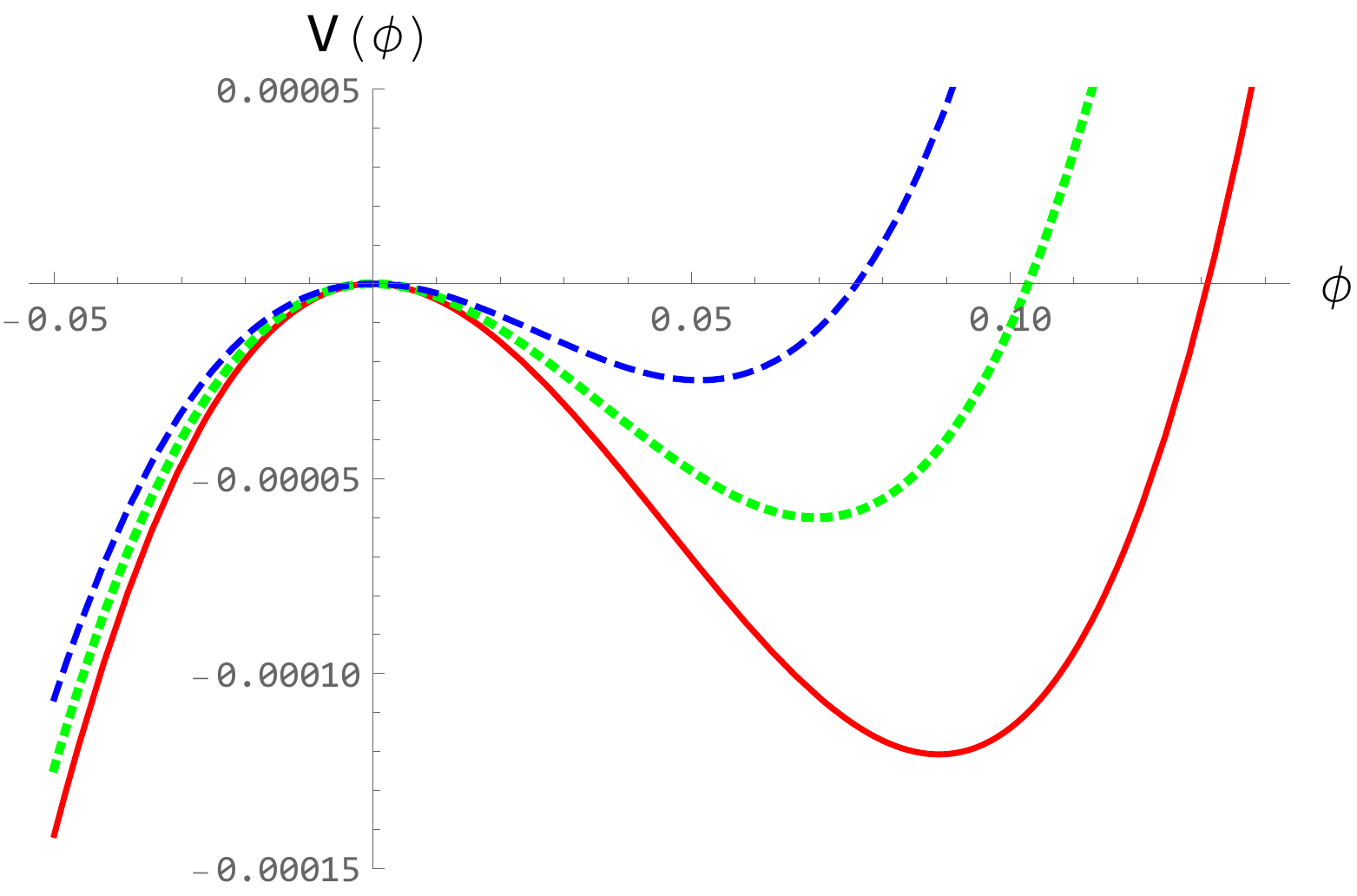}
\caption{The formation of potential wells in positive $\phi$-axis for $\sigma_{lt} = 0.02$ (solid curve), $\sigma_{lt} = 0.03$ (dotted curve), and $\sigma_{lt} = 0.04$ (dashed curve) at $\gamma_e = 5/3$, $\mathcal{M} = 1.21$, $\mu = 0.8$, $S_h = 0.5$, and $\sigma_{ht} = 0.015$. }
\label{Fig14}
\end{figure}
Note that we have used the same set of plasma parameters for both large and small amplitude limit. The results are shown in Figs. \ref{Fig9} - \ref{Fig11} (Figs. \ref{Fig12} - \ref{Fig14}) for small (large) amplitude NA SWs. It is observed that the effect of nucleus temperature reduces the possibility for the existence of subsonic NA SWs. Therefore, more number density or charge of heavy nucleus is required to have the subsonic NA SWs as the nucleus temperature rises. The amplitude (width) of both subsonic and supersonic NA SWs decreases (increases) with the increases in nucleus temperature. We have observed the same behaviour, as found in previous situation, that the ultra-relativistically DES ($\gamma_e = 4/3$) and non-relativistically DES ($\gamma_e = 5/3$) do not support the formation of subsonic NA SWs, but the BDES ($\gamma_e = 1$) does. The much wider NA solitary pulses in non-relativistically DES ($\gamma_e = 5/3$) as compared to the $\gamma_e = 1$ and $\gamma_e = 4/3$ have also been observed here.
\vspace{-8mm}
\section{\label{discus}Discussion}
The thermally degenerate pressure driven arbitrary amplitude nucleus acoustic solitary waves in a thermally degenerate plasma system (containing thermally degenerate electron species and non-degenerate light and heavy nucleus  species) have been investigated. The dynamics of light and heavy nucleus species has been studied based on equal footing. So, the solitary waves we investigated can be either light nucleus-acoustic solitary waves if $n_{l0}m_l\gg n_{h0}m_h$ or heavy nucleus-acoustic solitary waves if $n_{l0}m_l\ll n_{h0}m_h$. The pseudo-potential approach, which is valid for arbitrary amplitude solitary waves, has been employed.  The results, which have been obtained from this theoretical investigation, can be pinpointed as follows:
\begin{itemize}
\item{The phase speed of the  thermally degenerate nucleus-acoustic waves decreases (increases) with rise of the value of $\mu$ ($S_h$). The rate of decrease of the phase speed with $\mu$ in the case of $S_h\ne 0$ is slower than that in the case of $S_h=0$. This is due to same planarity of both dynamical species. However, the result would be opposite if the planarity of two dynamical species would be opposite.}

\item{It  is obvious that $C_q>C_l$  and $\lambda_q>L_q$. This means that the phase speed (wavelength) for $\mathcal{E}_{et} \ne 0$ is higher (lower) than that for $\mathcal{E}_{et}= 0$. This is due to the rise of the volume of the degenerate medium caused by the additional outward thermal pressure of the thermally degenerate electron species.}

\item{The consideration of Boltzmann distributed electron species ($\gamma_e=1$) makes the plasma system non-degenerate and gives rise to subsonic
thermally degenerate nucleus-acoustic solitary waves with $\phi>0$. However, the electron degeneracy and light and heavy nucleus temperature reduce the possibility for the formation of these supersonic solitary waves.}

\item{The Mach number decreases as the charge density of the heavy nucleus species increases which agrees with our linear analysis presented in introduction section. We note that the Mach number defined here is only valid if $n_{l0}m_l\gg n_{h0}m_h$ or if the waves are formed due to the compression and rarefaction of light nucleus species.}

\item{The consideration of ultra-relativistically and non-relativistically degenerate electron species supports  only the existence of supersonic solitary waves with $\phi>0$.}

\item{The amplitude (width) of both the subsonic and supersonic solitary waves decreases (increases) with the rise of values of $\gamma_e$,  $\sigma_{lt}$, and $\sigma_{ht}$. This is due to the fact that the latter increases the random motion of both light and heavy nucleus species.}

\item{The height of the solitary structures in non-relativistically degenerate electron species ($\gamma_e = 5/3$) is much smaller than that in ultra-relativistically degenerate electron species
($\gamma_e = 4/3$), but is much larger than that in Boltzmann distributed electron species ($\gamma_e = 1$).}

\item{The basic features obtained from analytical solitary wave solution of the energy integral with $V(\phi)=C_2\phi^2+C_3\phi^3$, which is valid for small but finite amplitude solitary waves,
are found to be the same as those obtained from the direct numerical analysis of the general form of $V(\phi)$, which is valid for arbitrary amplitude solitary waves.
This means that the basic features of the solitary waves identified in this investigations are correct.}
\end{itemize}
The electron-helium-carbon thermal degenerate plasma system (for which $Z_l = 2$, $Z_h = 6$, $m_e = 9.1 \times 10^{-31}$ kg, and $m_l = 1.6726 \times 10^{-27}$ kg,
$m_h = 2.0085 \times 10^{-26}$ kg)  have been used in our numerical analyses.  The  wide range of values of other parameters, viz. $\sigma_{lt} =0$-$0.3$, $\sigma_{ht} =0$-$0.3$,
and $\mu = 0.01$-$1$ have been used. Thus, the results obtained from this investigation are applicable in understanding the  salient features of localized electrostatic disturbances not only
in astrophysical compact objects like hot white dwarfs \cite{Dufour08, Dufour11,Werner15,Werner19,Koester20},  but also in space environments \cite{Rosenberg95,Havnes96,Tsintikidis96,Gelinas98} and laboratory devices \cite{Fortov96,Fortov98,Mamun08, Mamun09} where the electrons species follow the Boltzmann relation, the  ion species play the role as the light nucleus species does, and  the positively charged particles (as positively charged impurity or dust) play the role as the heavy nucleus species does.
\vspace{-4mm}
\acknowledgements
The authors are very  grateful to the  Alexander von Humboldt (AvH) Foundation (Bonn, Germany) for awarding  AvH Post-Doctoral Research Fellowship to A. Mannan, AvH Return Fellowship  to S. Sultana (after completing her AvH Postdoctoral Research Fellowship), and AvH Bessel Research Award to A A Mamun.

\end{document}